\documentclass[aps,prx,twocolumn,notitlepage,showpacs,amsmath,amstex,amssymb,citeautoscript,superscriptaddress,longbibliography]{revtex4-2}
\usepackage{graphicx}
\usepackage{hyperref}
\hypersetup{
    bookmarks=false,         % show bookmarks bar?
    unicode=false,          % non-Latin characters in AcrobatÕs bookmarks
    pdftoolbar=false,        % show AcrobatÕs toolbar?
    pdfmenubar=true,        % show AcrobatÕs menu?
    pdffitwindow=false,     % window fit to page when opened
    pdfstartview={FitH},    % fits the width of the page to the window
    pdftitle={},    % title
    pdfauthor={Ghazaryan et al},     % author
    pdfsubject={Multilayer graphenes as a platform for interaction-driven physics and topological superconductivity},   % subject of the document
    pdfcreator={},   % creator of the document
    pdfproducer={}, % producer of the document
    pdfkeywords={}, % list of keywords
    pdfnewwindow=true,      % links in new window
    colorlinks=true,       % false: boxed links; true: colored links
    linkcolor=black,          % color of internal links (change box color with linkbordercolor)
    citecolor=blue,        % color of links to bibliography
    filecolor=magenta,      % color of file links
    urlcolor=blue           % color of external links
}

\begin{document}

\title{Multilayer graphenes as a platform for interaction-driven physics and topological superconductivity}
\author{Areg Ghazaryan}
\affiliation{Institue of Science and Technology Austria, Am Campus 1, 3400 Klosterneuburg, Austria}
\author{Tobias Holder}
\affiliation{Department of Condensed Matter Physics, Weizmann Institute of Science, Rehovot 76100, Israel}
\author{Erez Berg}
\affiliation{Department of Condensed Matter Physics, Weizmann Institute of Science, Rehovot 76100, Israel}
\author{Maksym Serbyn}
\affiliation{Institue of Science and Technology Austria, Am Campus 1, 3400 Klosterneuburg, Austria}
\date{\today}
\begin{abstract}
Motivated by the recent discoveries of superconductivity in bilayer and trilayer graphene, 
we theoretically investigate superconductivity and other interaction-driven phases in multilayer graphene stacks. 
To this end, we study the density of states of multilayer graphene with up to four layers at the single-particle band structure level in the presence of a transverse electric field. 
Among the considered structures, tetralayer graphene with rhombohedral (ABCA) stacking reaches the highest density of states. 
We study the phases that can arise in ABCA graphene by tuning the carrier density and transverse electric field. 
For a broad region of the tuning parameters, the presence of strong Coulomb repulsion leads to a spontaneous spin and valley symmetry breaking via Stoner transitions.
Using a model that incorporates the spontaneous spin and valley polarization, we explore the Kohn-Luttinger mechanism for superconductivity driven by repulsive Coulomb interactions. 
We find that the strongest superconducting instability is in the $p-$wave channel, and occurs in proximity to the onset of Stoner transitions. 
Interestingly, we find a range of densities and transverse electric fields where superconductivity develops out of a strongly corrugated, singly-connected Fermi surface in each valley, leading to a topologically non-trivial chiral $p+ip$ superconducting state with an even number of co-propagating chiral Majorana edge modes.
Our work establishes ABCA stacked tetralayer graphene as a promising platform for observing strongly correlated physics and topological superconductivity.
\end{abstract}

\maketitle
Graphene heterostructures in presence of a superlattice potential induced by twist have opened new avenues to study interaction-driven physics~\cite{cao2018correlated,cao2018unconventional,balents2020superconductivity}. 
In addition, superconductivity was discovered in a variety of twisted materials~\cite{cao2018unconventional,park2021tunable,ParkFamily2021}, whose mechanism remains debated. 
In a parallel development, studies of rhombohedral trilayer graphene \emph{without} a moir\'e superlattice recently revealed interaction-driven ferromagnetic transitions~\cite{Zhou2021half} and superconductivity \cite{Zhou2021superconductivity}. 
Superconductivity has also been identified in Bernal-stacked bilayer graphene~\cite{ZhouBLG,NadjPerge2022}. The superconducting state is unconventional, at least in the sense that it far exceeds  the Pauli limit for an in-plane magnetic field.
These discoveries call for a detailed study of multilayer graphenes from the viewpoint of realizing interaction-driven symmetry broken phases, including superconductivity, which is addressed in the present work.

Conventional graphite is composed of graphene layers arranged in the so-called Bernal stacking, ABAB\ldots, where A and B denote two inequivalent graphene monolayers that are stacked in transverse direction. 
The study of interaction-driven physics in the bilayer AB graphene ---  one of the most-studied two-dimensional materials --- has a long history \cite{Mccann2013electronic,castro2008low,wang2007coulomb,vafek2010interacting, zhang2010spontaneous,jung2011lattice}. 
Its immediate relatives, trilayer ABA~\cite{ABA09,ABA11,ABA11T,ABA13,ABA16} and  tetralayer ABAB~\cite{Grushina15,Wu15,Shi18,Shi20} graphenes received considerably less attention to date. 
Moreover, the observation of interaction driven phenomena in trilayer and tetralayer Bernal stacks remains mostly limited to the regime of strong magnetic fields~\cite{ABA13} or suspended samples~\cite{Grushina15,ABAB18}.

In contrast to Bernal stacking, the so-called rhombohedral stacking is less stable, and thus less common. 
The simplest rhombohedral representative is ABC graphene that displays much richer interaction physics~\cite{ABC14,Zhou2021half} and even superconductivity when subjected to a perpendicular electric field~\cite{Zhou2021superconductivity}. 
Beyond three layers, transport~\cite{Myhro_2018} and scanning tunneling microscopy~\cite{ABCA21} studies in ABCA graphene revealed the development of a large gap at the neutrality point. 
Finally, thicker rhombohedral stacks were considered~\cite{ABC19,ABC20}, also revealing important interaction effects near charge neutrality. 

The dominant role of interaction effects in rhombohedral graphene, as compared to its Bernal allotrope, is naturally explained by a much higher density of states in the rhombohedral case. 
In particular, if one ignores further-neighbor hopping in the tight-binding model, the rhombohedral stacks of $n$ layers have an energy dispersion that depends on crystal momentum $k$ away from the corner of the Brillouin zone as $\pm|k|^n$, leading to a diverging density of states at charge neutrality for $n>2$~\cite{min2008electronic}. 
A transverse electric field that breaks inversion symmetry gaps out the band touching, leading to an even flatter dispersion that scales as $k^{2n}$ near the bottom of the conduction band and top of the valence band. 
A gap may also open by interactions at the neutrality point~\cite{zhang2011spontaneous,otani2010intrinsic}.
However, already for ABC trilayer graphene, the aformentioned approximation ignores  particle-hole asymmetry, trigonal warping, and other details of the band structure that are known to be important from experiments and from density functional theory calculations~\cite{otani2010intrinsic,pamuk2017magnetic}.  

This highlights the need for a theory of interaction effects in multilayer graphene, which incorporates the realistic band structure, extending the theory developed for bilayer and ABC trilayer graphene~\cite{Chou2021,Ghazaryan2021unconventional,Zale21,Levi21,MacDonaldABC22,Vish22,Guine22,Roy22,ABC-QMC,Chou2022acoustic}.  The rich physics of bilayer and ABC graphene away from neutrality point suggests multilayer graphene stacks as promising candidates for realization of symmetry broken phases and superconductivity. 

In this work we seek to identify the most promising multilayer graphene stacks for realizing correlated physics and exotic superconductivity.
To this end, we systematically study the noninteracting density of states of different multilayer structures. 
We identify the ABCA stacks of tetralayer graphene as the most promising candidate with the highest available density of states, that is further enhanced by the transverse electric field. 
For larger number of layers, we argue that stronger screening of the electric field precludes any further enhancement of the noninteracting density of states.
Therefore, in the remainder of our work we focus our attention on ABCA graphene. 

\begin{figure*}[t]
\centering
\includegraphics[width=.99\linewidth]{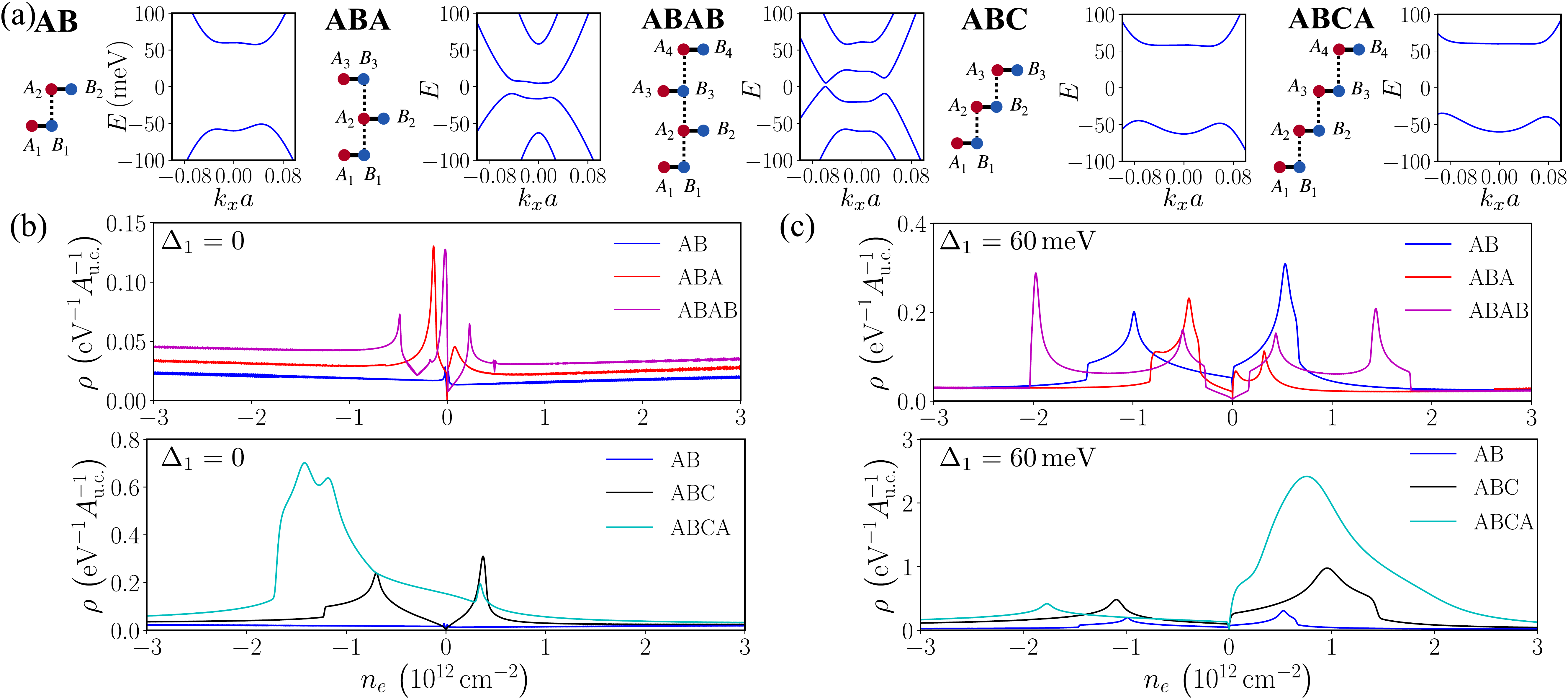}
\caption{
(a) Stacking order and corresponding energy dispersions for multilayers. The displacement field is $\Delta_1=60\,\mathrm{meV}$. 
The DOS as a function of the charge density is shown for $\Delta_1=0\,\mathrm{meV}$ (b) and $\Delta_1=60\,\mathrm{meV}$ (c).  Comparing the top plots in panels (b) and (c) reveals that among Bernal stacks AB graphene has broad regions with an enhanced DOS, for large $\Delta_1$. Bottom plots in (b) and (c) show that the DOS of ABC and ABCA graphene surpasses AB graphene by almost an order of magnitude. 
}
\label{fig:DOSCompareCombined}
\end{figure*}

ABCA graphene exhibits a rich fermiology as a function of density and transverse electric field with a variety of distinct Fermi surface types. 
This complex fermiology is due to the trigonal warping and particle-hole asymmetry which we include in our band structure model. 
We then consider the effect of interactions. 
Using a Stoner model, we identify regions of density and transverse electric fields that favor spin and valley symmetry broken phases. 
Finally, we study the leading superconducting instabilities within the Kohn-Luttinger scenario of superconductivity~\cite{KohnLuttinger,maiti2013superconductivity, kagan2015anomalous,Raghu2010,Chubukov2017}. 
Within this approach we identify the most promising regions of the phase diagram where the critical temperature of the superconducting instability may be accessible experimentally. 
Importantly, for a certain range of parameters a superconducting instability with $p-$wave pairing appears for a singly-connected Fermi surface in each valley, thereby realizing a topological superconducting phase in ABCA tetralayer graphene. 
The topological superconductivity theoretically predicted in our work does not require strong doping~\cite{Nandkishore2012chiral}, and could be experimentally verified using transport and tunneling measurements.  

We note that superconductivity in the surface states of bulk rhombohedral graphite was considered in Refs.~\cite{kopnin2011high,kopnin2013high}.  
These works assumed the presence of attractive interactions and demonstrated the enhancement of the critical temperature due to the flat-band character of the surface states in bulk graphite. 
In contrast, here we consider quasi-two dimensional systems with only few layers, which allows the application of a transverse electric field so that the carrier density can be changed by gating. Thus, we do not operate near the neutrality point, instead considering a finite carrier density and strong inversion-breaking electric fields. Moreover, we rely on the strong Coulomb repulsion that first gives rises to symmetry broken phases via Stoner transitions, and at the same time acts as a pairing glue within the Kohn-Luttinger mechanism. 

\section{Survey of multi-layer graphene band structures and density of states}
We begin with a survey of band structures and density of states (DOS) of graphene multilayers with different stackings. We consider systems up to four layers. Adding more layers does not introduce qualitative changes compared to tetralayers and may prevent control of layer potentials by gating in experiments due to the enhanced screening (see Appendix). The considered multilayers include Bernal stacked bilayer (AB), trilayer (ABA), tetralayer (ABAB) and corresponding rhombohedral stacked allotropes (ABC, ABCA), see Fig.~\ref{fig:DOSCompareCombined}(a). In addition, in the Appendix we consider the mixed stacking configuration ABCB which was recently realized experimentally~\cite{ABCB22}. The corresponding Hamiltonians are written in the basis $\left(A_1,B_1,A_2,B_2\dots\right)$, the size of the Bloch Hamiltonian matrix is thus $2n\times2n$, where $n$ is the number of layers, see Methods. Stackings AB, ABC, ABAB, and ABCA feature an inversion center, while ABA graphene has a mirror symmetry. We also consider the presence of a layer asymmetry potential due to the perpendicular electric field, $\Delta_1$ (defined such that the energy difference between outer layers is $2\Delta_1$, and the energy varies linearly with the layer index). 

\begin{figure}[t]
\centering
\includegraphics[width=.99\linewidth]{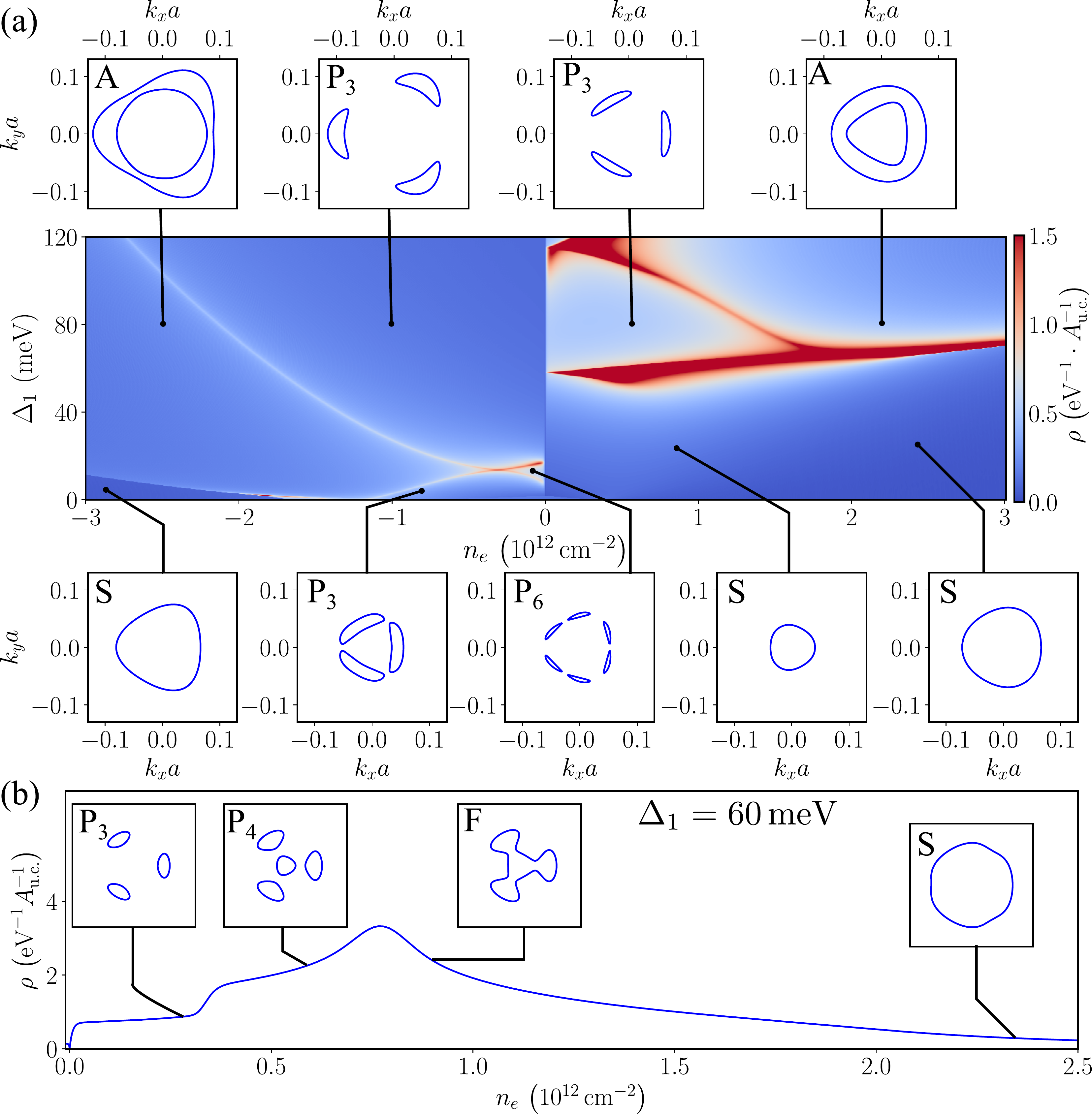}
\caption{(a) Color plot of the DOS $\rho$ for ABCA multilayer as a function of displacement field $\Delta_1$ and carrier density $n_e$. Different representative Fermi surface geometries are shown that are separated by VHS. Letters next to the Fermi surfaces designate different topologies, including single Fermi surface (S), three pockets (P$_3$), four pockets that include a central FS (P$_4$), six pockets (P$_{6}$), and annular geometry consisting of two concentric Fermi surfaces (A).
(b) Cut of the density of states for $n_e>0$ at fixed value of $\Delta_1 = 60\,{\rm meV}$ that reveals the existence of a single FS that is severely corrugated taking a ``flower'' shape (F). 
}
\label{fig:DOSFS}
\end{figure}
Examples of the low-energy band structures along the $k_x$ direction are shown in Fig.~\ref{fig:DOSCompareCombined}(a) for a relatively strong, but experimentally feasible value of  $\Delta_1=60\,{\rm meV}$. 
Only the AB, ABC and ABCA stackings possess a gap proportional to the applied electric field at charge neutrality. 
Including the trigonal warping and particle-hole asymmetry leads to a rich low energy dispersion, with multiple Fermi surface topologies separated by Van Hove singularities (VHS). As an overall trend, the DOS increases with increasing layer number. 
It should be noted that the wave functions of the low energy bands of rhombohedral graphene stacks are mostly localized on the outer layers, intuitively corresponding to (hybridized) edge states~\cite{Peres06}. In the three-dimensional limit, rhombohedral graphite is gapless, possessing Dirac nodal lines~\cite{mcclure1969electron,dresselhaus2002intercalation,heikkila2011dimensional,heikkila2011flat}. Thus we expect that for thicker rhombohedral stacks the stronger screening and appearance of multiple low-energy bands.

Fig.~\ref{fig:DOSCompareCombined} (b,c) shows the dependence of the DOS on charge density for different stackings for $\Delta_1=0\,\mathrm{meV}$ and  $\Delta_1=60\,\mathrm{meV}$, respectively. The top plots in panels (b) and (c) show the DOS of Bernal stacks, revealing that AB bilayer graphene has the highest DOS at large displacement fields in the Bernal family. The bottom plots in Fig.~\ref{fig:DOSCompareCombined}(b)-(c) compare the AB stacking with the trilayer and tetralayer rhombohedral stacks, illustrating the greatly increased DOS of the latter. Notably, ABCA stacking features the largest DOS for a broad region in density, making it a promising candidate for interaction-driven physics. We will now explore its properties in detail.

\section{Band structure of ABCA graphene}
\label{sec:ABCABand}
To reveal the large number of Fermi surface topologies  and the rich structure of the DOS of ABCA, we plot DOS as a function of charge density $n_e$ and layer asymmetry $\Delta_1$ in Fig.~\ref{fig:DOSFS}(a). $\Delta_1$ is proportional to the applied electric field, but the precise form of the relation is determined by the screening in the system. In the Appendix we consider Hartree type screening and show that values of $\Delta_1=120\,\mathrm{meV}$ can be realized in experiment through the application of a displacement field strength of $\sim2\,\mathrm{V/nm}$. 

On the hole side, $n_e<0$, we observe in total of five different FS topologies (ignoring for simplicity the region with $\Delta_1\lesssim 2\,\mathrm{meV}$,  where FS topology is complex and not relevant for the current discussion).  
In particular, a VHS divides regions with Fermi surfaces of three pockets denoted as P$_{3}$, annular (A) and simple single FS (S) in each valley. 
Note that for higher density range $n_e>-1\times10^{12}\,\mathrm{cm^{-2}}$, increasing $\Delta_1$ only changes the direction of the three pockets, corresponding to $\pi$ rotation of the pockets around $\mathbf{k}=0$ point [see top and bottom $\mathrm{P}_3$ panels on the hole side of Fig.~\ref{fig:DOSFS}(a)]. This transitions is accomplished by having a six pocket FS (P$_6$) in the vicinity of the VHS. 
Overall the FS topology of ABCA on the hole side is similar to ABC trilayer graphene \cite{Zhou2021half} with a somewhat enhanced DOS. 
This similarity suggests that the Stoner ferromagnetism and superconductivity mediated by electron-electron interactions will be qualitatively similar in both systems~\cite{Zhou2021half,Ghazaryan2021unconventional}. 

The electron side ($n_e>0$) of the phase diagram of ABCA is richer and qualitatively differs from the case of the ABC graphene. 
For small values of $\Delta_1$, there is only a single trigonally warped FS (S). Notably, the direction of the trigonal distortion (warping) is reversed upon increasing the charge density. This reversal is illustrated by the two insets at the bottom right corner of Fig.~\ref{fig:DOSFS}(a), that both show a simple FS that increases in size and changes the distortion orientation.  
In the Appendix we present the effective $2\times2$ model of the Hamiltonian of ABCA graphene stacks that explains the flipping of the trigonal warping.
At the density where this flipping occurs, the FS possesses six fold symmetry.

For larger values of $\Delta_1$ exceeding $80\,\mathrm{meV}$, the FS transitions from three the pockets regime (P$_3$) to an annulus (A). 
This transition is similar to  the one observed in ABC trilayer~\cite{Zhou2021half}, but is associated with considerably higher values of the DOS. 
In the range of $\Delta_1=60-80\,\mathrm{meV}$ the diagram shows a higher order VHS. 
In Appendix we zoom into that region and show that the high DOS there can be attributed to the existence of several higher order VHSs located nearby in the parameter space. In order to highlight the complexity of FS topologies, we show the density dependence of the DOS at a fixed value $\Delta_1=60$\,meV in Fig.~\ref{fig:DOSFS}(b). 
For this $\Delta_1$ we observe a transition between P$_3$ and P$_4$ FSs, see inset of Fig.~\ref{fig:DOSFS}(b). 
Moreover, we observe the merging of four pockets into a  severely corrugated flower-shaped singly connected FS (F). Although such topology of the FS is continuously connected to the single contour FS (labeled as S) at larger densities, it will play an important role in the subsequent discussion, as it allows for the realization of topological superconductivity.  

\section{Stoner and superconducting instabilities in ABCA graphene}
After considering the non-interacting DOS of ABCA graphene, we address interaction effects. Inspired by the experimental results for ABC trilayer graphene~\cite{Zhou2021half}, we first address the emergence of Stoner ferromagnetism. After understanding the sequence of Stoner transitions we investigate the leading superconducting instability. 

\begin{figure}[b]
\centering
\includegraphics[width=.999\linewidth]{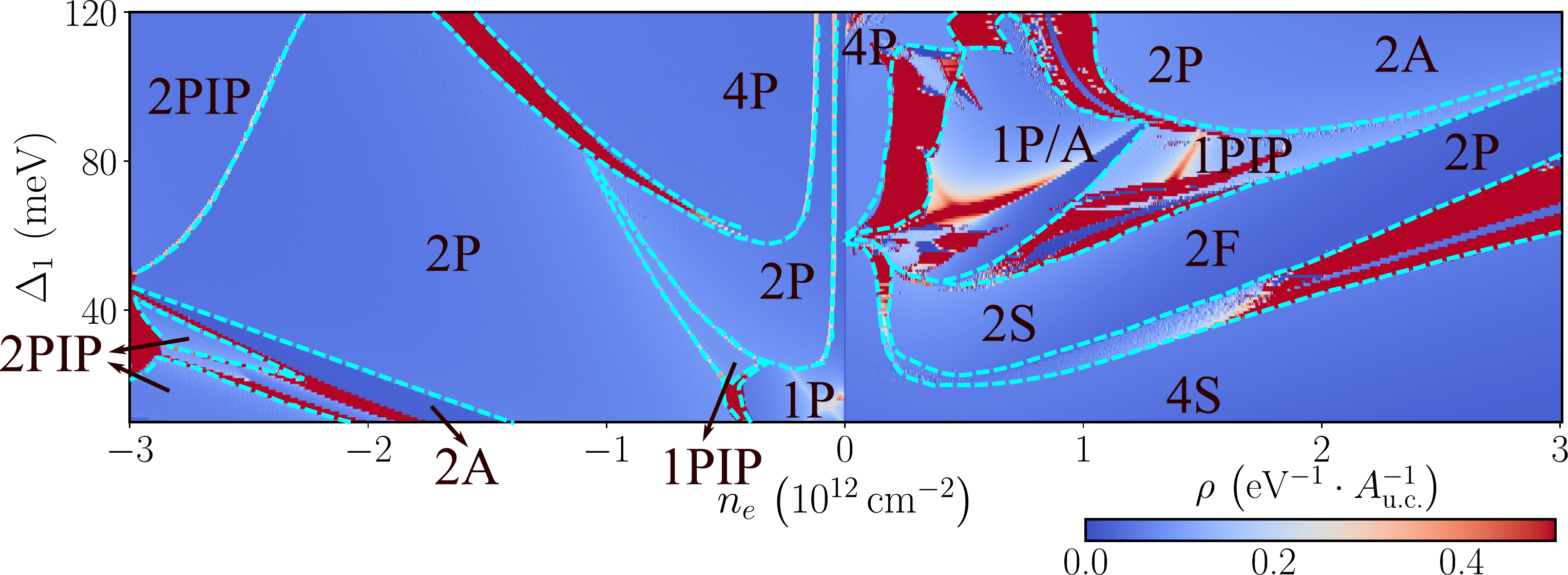}
\caption{Stoner phase diagram for interaction strengths $U=15\,\mathrm{eV}$ and $J=-4.5\,\mathrm{eV}$. Numbers denote the degeneracy of the phase and letters denote the topology of Fermi surfaces according to Fig.~\ref{fig:DOSFS}. For example, 2A is a region where twofold degenerate annular Fermi surface is realized. For the case of multiple pockets we use the label P without differentiating between different number of pockets (P$_3$, P$_4$, etc.). PIP denotes regions with partial isospin polarization. In the PIP regions we do not distinguish FS topologies. Different phases are separated by cyan dashed lines.}
\label{fig:Stoner}
\end{figure} 

\subsection{Stoner phase diagram}
We consider effects of the electron-electron interaction and the possibility of symmetry broken phases by means of a simple rigid band Stoner model which has been applied to twisted bilayer and ABC trilayer systems~\cite{Zondiner2020cascade,Zhou2021half}. 
The non-interacting bands are fourfold degenerate due to the spin and valley (denoted collectively as isospin) symmetry. 
Interaction effects can lead to spontaneous symmetry breaking, lowering the original fourfold flavor degeneracy.  
Assuming an SU(4)-symmetric form of the interaction, one naturally obtains phases with threefold degenerate, twofold degenerate, and non-degenerate bands. 
Experiments in ABC trilayer graphene \cite{Zhou2021half} did not reveal any regions with three-fold degenerate bands, a finding that can be attributed to the existence of lattice-scale interactions that break the SU(4) symmetry of the Coulomb interaction. 
We account for such terms phenomenologically, by adding a Hund's type contribution to the interactions that couples the spins of electrons from the two valleys (see Methods). 

The model thus includes two parameters for the interactions: $U$, which controls the strength of SU(4) symmetric part of the Coulomb repulsion, and the Hund’s coupling $J$, which sets the magnitude of terms breaking SU(4) symmetry.
Using this parametrization, it is possible to qualitatively capture the phase diagram of ABC trilayer graphene~\cite{Zhou2021half}, 
using values of  $U=30\,\mathrm{eV}$ and $J=-9\,\mathrm{eV}$. 
Expecting that interaction strength is weaker in ABCA tetralayer graphene, since adding more layers brings additional energy bands closer to neutrality point and enhances screening, 
we use $U=15\,\mathrm{eV}$ and $J=-4.5\,\mathrm{eV}$.  
Our choice of $J<0$ implies that the system prefers ferromagnetic ordering. This means that a doubly degenerate phase will be spin polarized and valley unpolarized as was observed in the ABC trilayer~\cite{Zhou2021half}. 

Figure~\ref{fig:Stoner} reveals the complex phase diagram obtained from the rigid Stoner model. Due to the considerable magnitude of the Hund's ferromagnetic term, we do not observe phases with three-fold degeneracy. 
At low values of $\Delta_1$, we do not observe any symmetry broken phases in the electron-doped side, whereas on the hole-doped side ($n_e<0$) we see the sequence 4P, 2PIP, 2A, 2P, 1P of symmetry broken phases, where the numbers denote the degeneracy of the phase, while letters denote the FS topology (the notation follows Fig.~\ref{fig:DOSFS}, except for the case of multiple pockets in which case we do not differentiate between different number of pockets and use the general label P). 
PIP corresponds to partially isospin polarized phase, where, for example, for 2PIP all four flavors are filled, but two have smaller filling compared to the remaining two. 
For PIP phases we do not differentiate between phases with distinct Fermi surface topologies. 
Upon increasing the perpendicular electric field,  extended regions of two-fold degenerate phases develop on the electron side with multiple pockets, simple or an annular FS. 
Moreover in the parts of 2S region for positive densities ($n_e\approx 1.5 \cdot 10^{12}\,{\rm cm}^{-2}$), the flower-type geometry of FS is realized.
This will have important consequences on superconducting instabilities for the charge densities and asymmetry potential values where these phases are realized. At even higher values of $\Delta_1$ the 2A region prevails, while at lower densities one also observes region of single-degenerate FS (1P/A).

Finally, let us comment on the choice of the interaction parameters $U$ and $J$ and their effect on the phase diagram. 
In the Appendix we present results for the phase diagram with parameters identical to the ABC interaction strength used in~\cite{Zhou2021half}.
In that latter scenario, the significantly larger interaction strength suppresses the PIP phases in the phase diagram. In addition, 
since the spontaneous symmetry breaking is most prominent close to band edges and lower densities, the increased interaction strength results in stronger symmetry breaking in the regime of small densities. 
Generically, the increased prevalence of symmetry broken phases suppresses superconductivity, since the growth of the DOS in the normal state (conducive to superconductivity) is preempted by a Stoner instability.

\subsection{Leading superconducting instabilities}
 We consider the potential superconducting instabilities of ABCA graphene using the band structure as obtained from the Stoner instability analysis.  To this end, it is henceforth assumed that pairing emerges due to electron-electron interactions by the Kohn-Luttinger mechanism~\cite{KohnLuttinger,maiti2013superconductivity,kagan2015anomalous}. In this mechanism attractive interaction between electrons is mediated through electron-hole fluctuations of electronic fluid. We account for these fluctuations through the random phase approximation (RPA), as in Ref.~\cite{Ghazaryan2021unconventional}. We use the Fermi surfaces and structure of non-interacting wave functions obtained earlier for the Fermi surfaces with broken degeneracy due to Stoner mechanism. The parameter $N=4$ (no broken symmetry) or $2$ (twofold degenerate bands) controls the number of occupied flavors, thereby affecting the screening. We assume that phases with $N=1$ are valley polarized, and thus have no superconducting instability. As was noted in Sec.~\ref{sec:ABCABand}, on the hole side the non-interacting phase diagram is qualitatively similar to the hole side in ABC trilayer graphene. Therefore, here we concentrate on the electron side of the phase diagram and on the new superconducting phases that were not present in ABC trilayers. 
 
 After obtaining the effective electron-electron interaction within RPA, $V_\mathbf{q}$, we compute the superconducting coupling constant $\lambda$ from the linearized BCS gap equation. In the weak coupling approximation, $\lambda$ determines the superconducting critical temperature according to $T_c = We^{-1/\lambda}$ with $W$ being an energy cutoff, typically of order of the Fermi energy. Alongside $\lambda$, we also study the gap function to infer the symmetry of the order parameter of the superconducting phase. 
 
\begin{figure}[t]
\centering
\includegraphics[width=.99\linewidth]{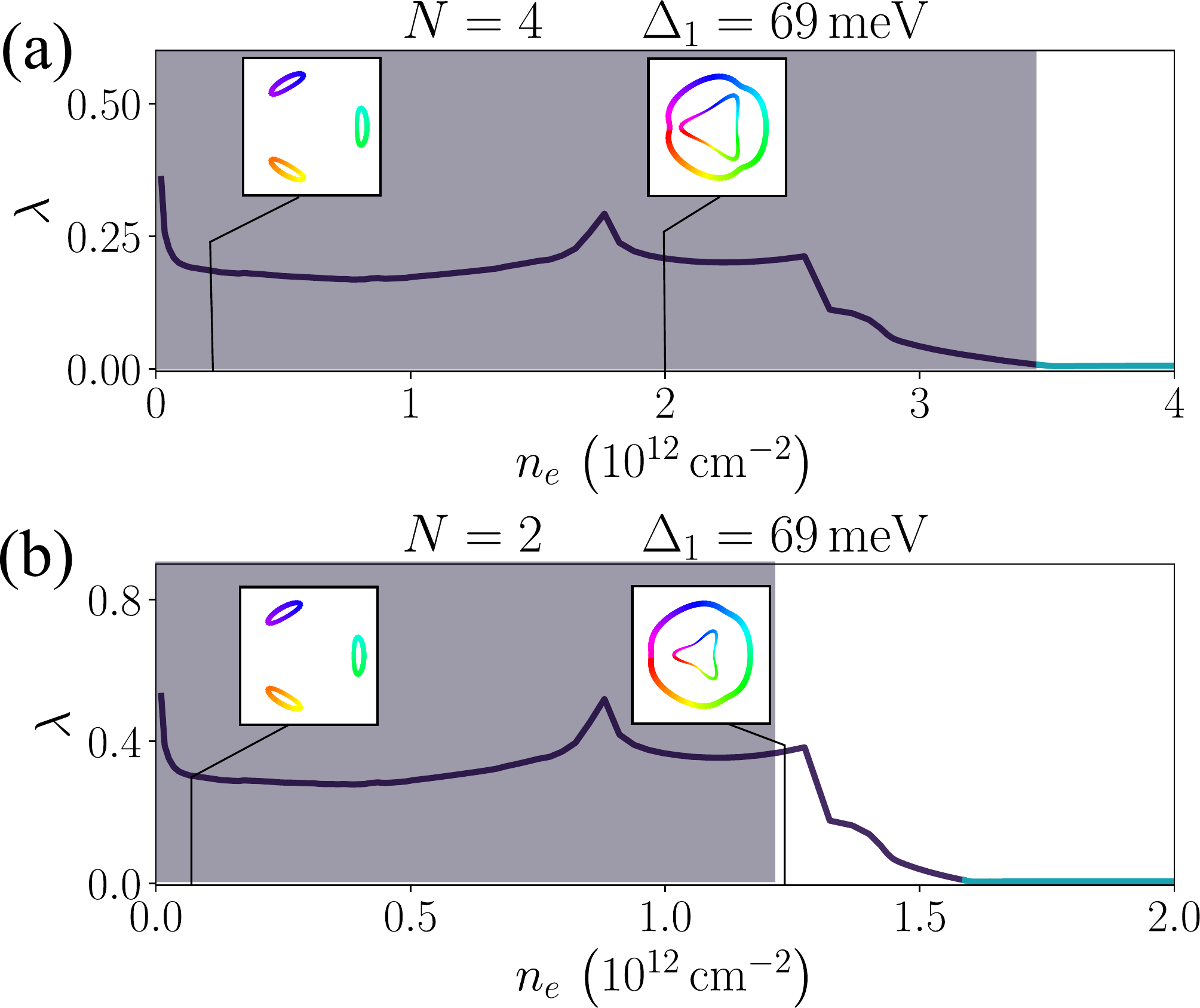}
\caption{Dimensionless superconducting coupling constant $\lambda$ in ABCA tetralayer graphene as a functon of density for $\Delta_1=69\,\mathrm{meV}$, assuming a degeneracy $N=4$ for panel (a) and $N=2$ for panel (b).  
In both cases, there is a very weak instability towards pairing in a high angular momentum channel at high densities (cyan color), 
followed by a much stronger $p-$wave pairing instability at lower densities (violet color).  
Shaded regions correspond to the range of densities that are inaccessible at the given degeneracy due to occurrence of the Stoner phase transition which decreases the degeneracy and changes the Fermi surface. Insets show Fermi surfaces and color denotes the phase of the order parameter, visually representing the chiral nature of the state.}
\label{fig:SuperStand}
\end{figure} 

First we investigate the behavior of the coupling constant at a relatively high value of $\Delta_1$. In that case,  the FS geometry changes from P$_3$ to annular. 
The presence of two FS contours results in considerable enhancement of superconductivity due to the Kohn-Luttinger mechanism~\cite{Raghu2010,Raghu2011,Chubukov2017}. Similar to the ABC trilayer case, we find the dominant instability to have $p$-wave symmetry, both for $N=4$ and $N=2$ (see Fig.~\ref{fig:SuperStand}). For a $C_3$ symmetric system, the quartic term in the free energy then favors a chiral $p+ip$ state below $T_c$~\cite{Ghazaryan2021unconventional}. In addition, we also find a small region with extended $s$-wave pairing (not shown).

Next, we combine the superconducting ordering tendencies with the Stoner phase diagram to identify regions where flavor degeneracy due to symmetry breaking is compatible with the normal phase of superconductor. As is clear from Fig.~\ref{fig:SuperStand}, for twofold degeneracy ($N=2$) there are indeed regions of large $\lambda$ which are not negatively by the Stoner instability. Notably, assuming that the $N=2$ state is spin polarized, we predict the realization of spin-triplet superconductivity with $p+ip$ symmetry of the order parameter. However, despite the presence of $p$-wave pairing, the superconducting phase is not topological. The reason for this being the two contours of the annular Fermi surface, which have opposite Chern numbers and overall add up to a vanishing Chern number. 
When comparing the results of ABCA tetralayer with ABC trilayer, the coupling constant $\lambda$ for ABCA graphene is considerably larger. 
As mentioned earlier, this is related to the enhanced DOS of ABCA. Therefore, we expect the superconducting instability to be more robust than for the ABC trilayer.

\begin{figure}[t]
\centering
\includegraphics[width=.99\linewidth]{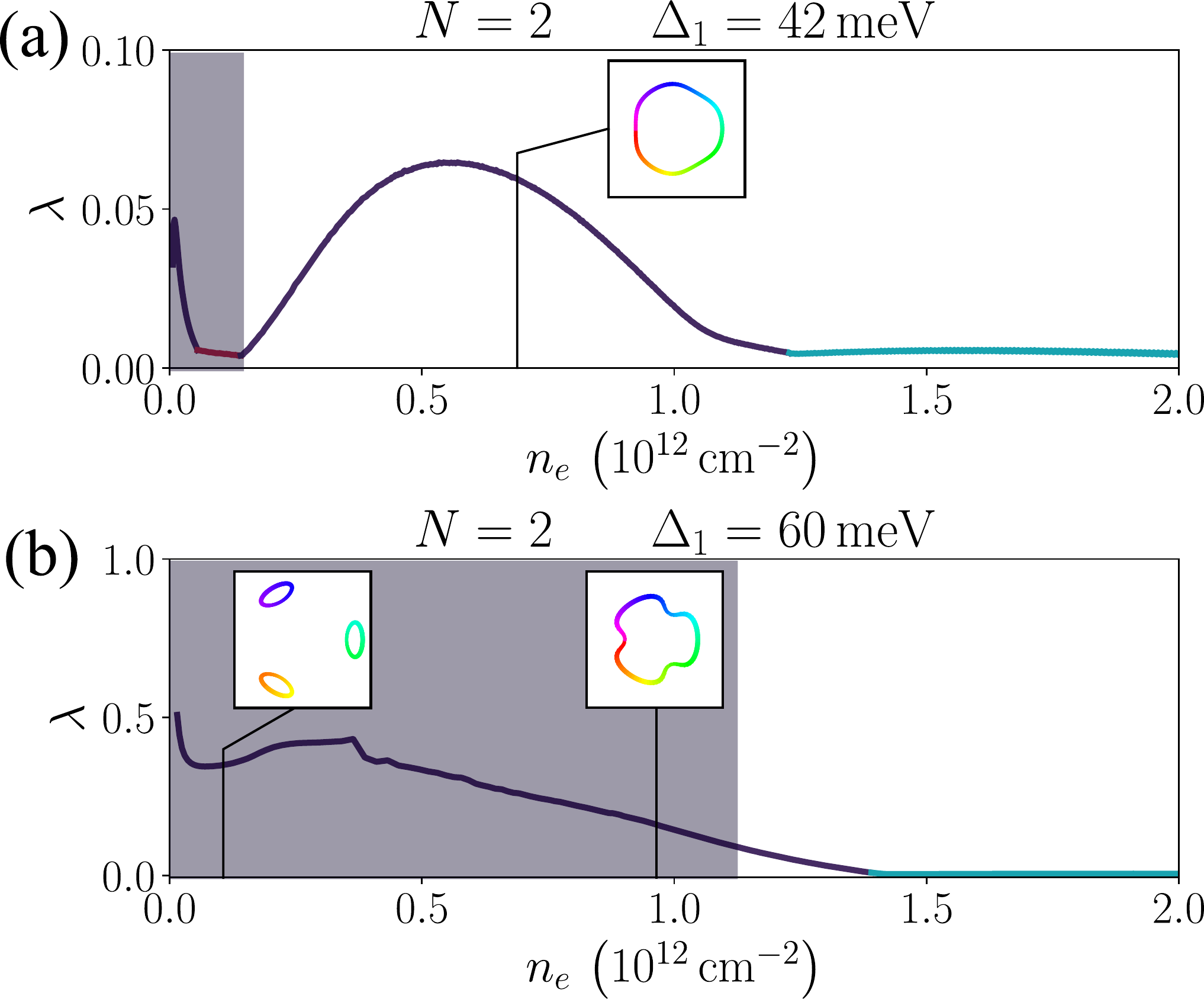}
\caption{Focusing on the superconducting instability in the regime of two-fold degenerate Fermi surfaces ($N=2$) reveals a range of densities with intermediate values of $\lambda \geq 0.05$ and a single, simply connected Fermi surface. This situation is realized for $\Delta_1=42\,\mathrm{meV}$ (a) and $\Delta_1=60\,\mathrm{meV}$ (b) and leads to topological superconductivity.}
\label{fig:SuperTop}
\end{figure} 

\subsection{Regime of topological superconductivity}
 For somewhat lower values of $\Delta_1$, when a single, simply-connected FS is present (see S and F contours in Fig.~\ref{fig:DOSFS}), topological superconductivity can be realized. To this end, we focus on the case when the Stoner transition reduces the FS degeneracy to $N=2$ and study the density dependence of $\lambda$ for two different values of $\Delta_1$. Figure~\ref{fig:SuperTop}(a) corresponds to value of $\Delta_1<45\,\mathrm{meV}$, when the FS contour is trigonally distorted and switches the direction of the trigonal distortion as a function of electron density. At the position where the direction of trigonal distortion is changed, the FS enjoys an approximate six-fold rotation symmetry, resembling a distorted hexagon. This shape of FS shows features of nesting, leading to enhancement of  static polarization at non-zero momenta, and also enhancing superconducting instability. The peak of DOS appears at smaller densities than the position where trigonal warping direction is flipped (not shown). Since the DOS also affects superconducting instability, the coupling constant and correspondingly the critical temperature shows a maximum at the density which is in between the densities of peak of DOS and ``hexagonal'' FS.
 The maximal values of $\lambda$ in this regime are lower than those obtained for the annular FS (cf. Fig.~\ref{fig:SuperStand}). Nevertheless, the enhancement of $\lambda$ suggests that enhancement of symmetry of FS due to change in trigonal warping direction gives rise to sizable critical temperatures within the framework of the Kohn-Luttinger approach, even in absence of a multi-pocket Fermi surface.
 
 The superconducting state illustrated in Fig.~\ref{fig:SuperTop}(a) is topologically non-trivial. For intermediate values of the charge density, the pairing is $p$-wave, and we expect the resulting order parameter to have $p+ip$ character. For $N=2$, there is a single FS in each valley. Each FS has Chern number of one, so in total the system has Chern number of two. Notably, for this value of $\Delta_1$, the Stoner instability that further reduces the valley degeneracy occurs at lower electron densities. Thus we predict a dome-shaped superconducting region as a function of carrier density, terminated by Stoner transition at low densities around $n_e\approx 0.15\cdot 10^{12}$\,cm$^{-2}$.
 
The $\lambda$ obtained for the topological superconductivity from a single, convex FS turned out to be relatively small (Fig.~\ref{fig:SuperTop}(a)). This changes closer to VHS where the single contour FS takes on a flower shape (F contour in Fig.~\ref{fig:DOSFS}). As shown in Fig.~\ref{fig:SuperTop}(b), for the flower-shaped FS, $\lambda$ grows rapidly with decreasing density. 
Thus, we observe that within the Kohn-Luttinger mechanism, the corrugated nature of the single-contour Fermi surface substantially enhances the superconducting coupling constant. 
Here the leading superconducting instability again has $p$-wave symmetry with non-zero Chern number (cf.~SI). While $\lambda$ rapidly increases with decreasing density, the appearance of Stoner transition at higher densities [shaded region in Fig.~\ref{fig:SuperTop}(b)] in this case precludes the development of dome shaped superconducting region. Therefore, for large values of $\Delta_1$ superconductivity is only limited to narrow region adjacent to Stoner instability similar to ABC trilayer. 

Overall, we find that the ABCA tetralayer is qualitatively different from ABC trilayer since it gives rise to regions with robust topological superconductivity. 
In particular, the Kohn-Luttinger mechanism suggests a $p+ip$ topological superconducting phase immediately preceding the Stoner transition at large perpendicular electric fields, $58\,\mathrm{meV}\leq\Delta_1\leq68\,\mathrm{meV}$. At even larger fields, the superconducting transition of ABCA tetralayer for the conduction band should be similar to ABC trilayer, with an annular-shaped Fermi surface and a topologically trivial superconducting state.

\section{Discussion}
We investigated multilayer graphenes as potential candidates for realization of symmetry broken phases and superconductivity, identifying tetralayer ABCA stacks of graphene as a the most promising candidate for realizing interaction-driven physics. 
The dominant superconducting instability has a $p-$wave order parameter and typically occurs for the annular or pocket geometry of the Fermi surface. This is in line with the established intuition that Kohn-Luttinger mechanism is enhanced in presence of multi-pocket Fermi surfaces~\cite{KohnLuttinger,Raghu2010}. According to our findings, the superconducting regions are expected to trail the Stoner transitions --- a qualitative prediction that is testable in future experiments. Somewhat surprisingly, we also found significant superconducting instabilities for certain simply connected Fermi surfaces that either have an approximate six-fold rotation symmetry or are severely corrugated (i.~e. the FS has a flower shape). We predict that the superconducting phase resulting from such simply connected Fermi surfaces is topological. 

The topological superconducting phase will manifest itself with topologically protected two co-propagating chiral edge modes originating from the two valleys. 
These edge states can be detected in tunneling experiments into the edges, or through their quantized contribution to the thermal Hall conductivity of $\kappa_{XY} = {\pi^2 k_B^2 C}/({6h})$~\cite{Read2000}, where $C$ is the Chern number of the superconducting state.

We note that a scenario for (non-topological) superconductivity due to phonons has recently been proposed for ABCA graphene~\cite{Chou2022}. In contrast to our results, the phonon mechanism predicts dome-shaped superconducting regions with $s$ or $f$-wave symmetry for a broad range of densities, and for large values of the electric field. While we do find some regions with a dome-shaped superconducting instability, in our analysis the most robust superconducting states occur in narrow density ranges near the border between two phases with different patterns of valley and spin symmetry breaking.  

Topological superconductivity from a single-connected Fermi surface is a distinct possibility that emerges only for tetralayer graphene. Conceptually, it suggests that the Kohn-Luttinger mechanism can give rise to a sizable instability also for the simply-connected Fermi surfaces, provided these are sufficiently far from a circular Fermi surface with parabolic dispersion. It would be interesting to study the Kohn-Luttinger scenario for other systems characterized by a strongly warped or distorted Fermi surface at strong Coulomb interactions. Independent of that, the approximate six-fold symmetry of the Fermi surface which we find in ABCA graphene is expected to occur in other multilayer two-dimensional materials with a hexagonal lattice. An accurate treatment of potential competing instabilities of such approximately-nested Fermi surface remains an interesting open question.

To conclude, we reveal that despite its relative complexity, tetralayer graphene holds the promise of realising new physics, including topological superconductivity, that was hitherto not observed in a graphene stacks with a smaller number of layers. Although we identified ABCA graphene as the most promising material, a large number of potential other systems remain beyond the scope of our work. This applies in particular to uncoventional ABCB stacks, which have a relatively high density of states as well (cf. SI) and which were recently realized experimentally~\cite{ABCB22}. In addition, our Stoner model did not include possible nematic phases that would break the three-fold rotation symmetry of underlying graphene lattice. These phases are likely to occur in the regime of low density, when the Fermi surface typically contains several small pockets~\cite{Jung2015}. We hope that joint future theoretical and experimental studies will advance our understanding of interaction effects of multilayer graphenes, thereby facilitating the realization of new phases of matter.

\section{Methods}
\subsection*{Band structure and tight binding parameters}
The non-interacting band structure of the ABCA graphene is derived from the eight band continuum model. Adopting the standard Slonczewski-Weiss-McClure parametrization of the tight-binding model the effective Hamiltonian can be written in the form
\begin{equation}
 H=H_{0}+H_{\Delta_1}+H_{\Delta_2}+H_{\Delta_3},   
\end{equation}
where the last three terms denote layer potentials emergent due to applied external field and non-trivial charge distribution among four graphene layers. Written in the basis of $\left(A_1,B_1,A_2,B_2,A_3,B_3,A_4,B_4\right)$, where $A_i$ and $B_i$ denote different sublattice sites on layer $i$, these terms read
\begin{align}
& H_{0}=\notag\\
& \left(\begin{array}{cccccccc}
0 &v_0\pi^\dagger &v_4\pi^\dagger &v_3\pi &0 &\frac{1}{2}\gamma_2 &0 &0  \\
v_0\pi &\delta &\gamma_1 &v_4\pi^\dagger &0 &0 &0 &0 \\
v_4\pi & \gamma_1 &\delta &v_0\pi^\dagger &v_4\pi^\dagger &v_3\pi &0 &\frac{1}{2}\gamma_2 \\
v_3\pi^\dagger &v_4\pi &v_0\pi &\delta &\gamma_1 &v_4\pi^\dagger &0 &0 \\
0 &0 &v_4\pi &\gamma_1 &\delta &v_0\pi^\dagger &v_4\pi^\dagger &v_3\pi \\
\frac{1}{2}\gamma_2  &0 &v_3\pi^\dagger &v_4\pi &v_0\pi &\delta &\gamma_1 &v_4\pi^\dagger \\
0 &0 &0 &0 &v_4\pi &\gamma_1 &\delta &v_0\pi^\dagger \\
0 &0 &\frac{1}{2}\gamma_2  &0 &v_3\pi^\dagger &v_4\pi &v_0\pi &0			\end{array}\right), 
\end{align}
\begin{align}
H_{\Delta_1}&=\mathrm{diag}\left(\Delta_1,\Delta_1,\frac{\Delta_1}{3},\frac{\Delta_1}{3},-\frac{\Delta_1}{3},-\frac{\Delta_1}{3},-\Delta_1,-\Delta_1\right),\\
H_{\Delta_2}&=\mathrm{diag}\left(\Delta_2,\Delta_2,-\Delta_2,-\Delta_2,-\Delta_2,-\Delta_2,\Delta_2,\Delta_2\right), \\
H_{\Delta_3}&=\mathrm{diag}\left(0,0,-\Delta_3,-\Delta_3,\Delta_3,\Delta_3,0,0\right).
\end{align}
Here we denote $\pi=\tau k_x+ik_y$, with $\tau=\pm1$ being the valley index. The band structure parameters $v_i=\sqrt{3}a\gamma_i/2$ for $i=0,3,4$, $a=2.46\AA$ is the lattice constant of graphene and $H_0$ depend on hopping matrix elements $\gamma_0\dots\gamma_4$ which correspond to effective hopping amplitudes up to three layers. $\delta$ is the onsite potential for the sublattice sites which have direct neighbor on adjacent layer and $\Delta_1$, $\Delta_2$ and $\Delta_3$ determine the electrostatic potential between the layers. For hopping and onsite potential parameters we use the values of ABC graphene~\cite{Zhou2021half}:\\[-15pt]
\begin{table}[h!]
\centering
\begin{tabular}{ccccccc}
\hline\hline
$\gamma_0\,(\mathrm{eV})$ & $\gamma_1\,(\mathrm{eV})$ &$\gamma_2\,(\mathrm{eV})$ & $\gamma_3\,(\mathrm{eV})$ &$\gamma_4\,(\mathrm{eV})$ &$\delta\,(\mathrm{eV})$  \\
\hline
$3.1$ & $0.38$ &  $-0.015$ & $-0.29$ & $-0.141$ & $0.0105$ \\
\hline\hline
\end{tabular}
\label{table1}
\end{table}

\noindent Hamiltonians for other stackings and other number of layers can be written in a similar fashion and are presented in the Appendix. 

\subsection*{Layer potentials and screening}
The electrostatic potential parameters $\Delta_1$, $\Delta_2$ and $\Delta_3$ for ABCA graphene are determined from individual layer potentials $u_i$ through 
\begin{align}
\Delta_1&=\frac{u_1-u_4}{2}, \\
\Delta_2&=\frac{u_1-u_2-u_3+u_4}{4}, \\
\Delta_3&=\frac{u_1-3u_2+3u_3-u_4}{6}.
\end{align}
From this definition, $2\Delta_1$ is the potential difference between outer layers, $2\Delta_2$ is the average potential difference between outer and inner layers, and $2\Delta_3$ denotes potential difference between inner layers when $\Delta_1=0$. From electrostatics it can be shown that $\Delta_1$ depends on external displacement field and layer charge densities, whereas $\Delta_2$ and $\Delta_3$ depend on layer charge densities of inner layers. Since layer charge densities also depend on electrostatic potential parameters, the potentials should be determined self-consistently for each value of external electric field. Employing Hartree approximation for screening we get values of $\Delta_2$ and $\Delta_3$ in the meV range for physically realistic values of displacement field. Therefore, their effect on the band structure is rather small and we fix $\Delta_2=\Delta_3 = 0$ in the calculations. For $\Delta_1$ we estimate the maximum value attainable in the experiment to be around $120\,\mathrm{meV}$ corresponding to the displacement field of $D\approx2.0\,\mathrm{V/nm}$. 

\subsection*{Simplified Stoner model}
To understand the nature of symmetry broken phases we employ a simplified Stoner model with four flavors. Denoting the flavors as $1=\{K,\uparrow\}$, $2=\{K^\prime,\uparrow\}$, $3=\{K,\downarrow\}$, $4=\{K^\prime,\downarrow\}$, where $K,K^\prime$ specify the valleys and $\uparrow,\downarrow$ the spin, the grand potential per area is defined as
\begin{equation}
\frac{\Phi}{A}=\sum_\alpha E_{0}\left(n_\alpha\right)+V_\mathrm{int}-\mu\sum_\alpha n_\alpha,    
\end{equation}
where $E_{0}(n_\alpha)$ is the kinetic energy calculated from non-interacting band structure for flavor $\alpha$ with the charge density $n_\alpha$ and $\mu$ is the chemical potential. The interaction potential $V_\mathrm{int}$ is taken as
\begin{equation}
V_\mathrm{int}=\frac{UA_\mathrm{u.c.}}{2} \sum_{\alpha\neq\beta}n_\alpha n_\beta+JA_\mathrm{u.c.}\left(n_1-n_3\right)\left(n_2-n_4\right).
\end{equation}
Here $1=\{K,\uparrow\}$, $2=\{K^\prime,\uparrow\}$, $3=\{K,\downarrow\}$, $4=\{K^\prime,\downarrow\}$ and $A_\mathrm{u.c.}=\sqrt{3}a^2/2$ is the area of unit cell, $U$ is the valley and spin-isotropic interaction constant and $J$ is inter-valley spin-exchange Hund's rule coupling constant, which explicitly breaks SU(4) symmetry. $J<0$ ($J>0$) favours valley-unpolarized ferromagnetic (antiferromagnetic) phase when only two flavours are occupied. To determine the realized phase for each chemical potential, we minimize the grand potential for interaction strengths  $U=15\,\mathrm{eV}$ and $J=-4.5\,\mathrm{eV}$.

\subsection*{Superconducting instability}
To obtain the pairing interaction we consider screened Coulomb interaction between the electrons $V_{0,\mathbf{q}}=\frac{2\pi e^{2}}{\epsilon q}\tanh(qd)$, where $\epsilon$ is the dielectric constant of ABCA graphene, $d$ is the distance to the metallic gates located on both sides of the sample. We use $\epsilon=4$ and $d=36.9\,\mathrm{nm}$. To incorporate electron-hole fluctuations we use random phase approximation (RPA)
\begin{equation}
V_{\mathbf{q}}=\frac{V_{0,\mathbf{q}}}{1+\Pi_{0,\mathbf{q}}V_{0,\mathbf{q}}},    
\end{equation}
with $\Pi_{0,\mathbf{q}}$ being the static polarization function. $\Pi_{0,\mathbf{q}}\propto N$ occupied flavours and this controls both the screening and strength of superconducting instability. Coupling constant $\lambda$ is obtained from diagonalizing the eigenvalue equation $\mathcal{M}\left(V_\mathbf{q}\right)\Delta_\mathbf{k}=\lambda\Delta_\mathbf{k}$, where $\mathcal{M}\left(V_\mathbf{q}\right)$ is a linear operator, which is proportional to vertex of scattering of a pair of electrons from one momentum to the other on a FS. The largest eigenvalue $\lambda$ corresponds to the leading superconducting instability, and the shape of the wave function may be used to infer the corresponding symmetry of the order parameter. For the case of $p-$wave symmetry the nature of order parameter below $T_c$ can be inferred from Ginzburg-Landau energy functional. Quite generally, when system possesses $C_3$ symmetry chiral superconducting state is preferred~\cite{Ghazaryan2021unconventional}.  
\begin{acknowledgments}
E.B. and T.H. were supported by the European Research Council (ERC) under grant HQMAT (Grant Agreement No.~817799), by the Israel-USA Binational Science Foundation (BSF), and by a Research grant from Irving and Cherna Moskowitz.
\end{acknowledgments}
\appendix
\section{Tight binding model for band structure and screening}

\begin{figure}[b]
\includegraphics[width=0.99\linewidth]{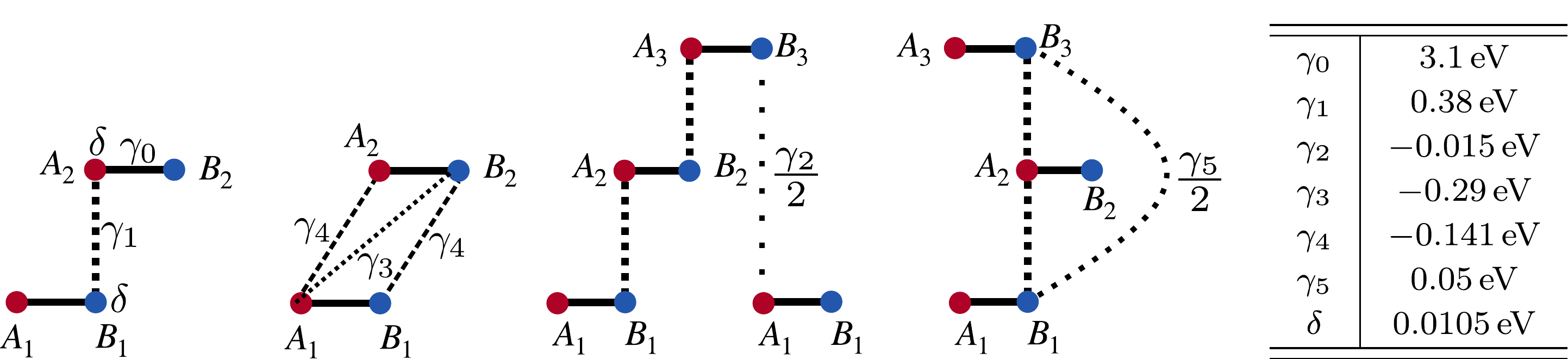}

\caption{Schematic representation of hopping and on site parameters $\gamma_i$ and $\delta$ (left) and corresponding values used in the current calculations (right). 
}
\label{fig:hoppingparam}
\end{figure}
\subsection{Bilayer and trilayer graphenes}

To consider band structures of multilayer graphene we adopt the standard Slonczewski-Weiss-McClure parametrization of the tight-binding model~\cite{Dresselhaus1981intercalation}. We retain hopping amplitudes effective up to three layers, as shown in Fig.~\ref{fig:hoppingparam}. The parameter values are adopted from Ref.~\cite{Zibrov2018emergent,Zhou2021half}. In particular, since we are interested in ABCA stacking, we use the parameters proposed for ABC graphene \cite{Zibrov2018emergent}. For certain stackings, an additional hopping $\gamma_5$ is present (see Fig.~\ref{fig:hoppingparam}), which is absent for ABC graphene. In this case we adopt the value estimated for ABA graphene \cite{Zibrov2018emergent}. The band structure of bilayer and trilayer graphene has been considered before \cite{Mccann2013electronic,Jung2014accurate,MacDonald2010,Koshino2009,Koshino2010,Zhou2021half}.  For completeness we reproduce here the tight-binding Hamiltonian of the bilayer and trilayer. Denoting $A_i$ and $B_i$ for the different sublattice sites on layer $i$, for the bilayer the basis is $\left(A_1,B_1,A_2,B_2\right)$ and we write
\begin{equation}
H_\mathrm{BG}=\left(\begin{array}{cccc}
 \Delta_1 &v_0\pi^\dagger &v_4\pi^\dagger &v_3\pi \\
 v_0\pi &\Delta_1+\delta &\gamma_1 &v_4\pi^\dagger \\
 v_4\pi & \gamma_1 &-\Delta_1+\delta &v_0\pi^\dagger \\
 v_3\pi^\dagger &v_4\pi &v_0\pi &-\Delta_1
\end{array}\right),
\end{equation}  
where $\pi=\tau k_x+ik_y$ ($\tau$ is the valley index) and $v_i=\sqrt{3}a\gamma_i/2$, $a=2.46\AA$ is the lattice constant of graphene. $2\Delta_1$ is potential difference between the layers and is proportional to the perpendicular electric field and $\delta$ is the onsite potential for the sublattice sites which have a direct neighbor on the adjacent layer (see Fig.~\ref{fig:hoppingparam}).
For the trilayer case we have two stackings $\mathrm{st}=\mathrm{ABA}$ and ABC, so we use the following parametrization
\begin{equation}
H_{\mathrm{st}}=H_{\mathrm{st},0}+H_{\Delta_1}+H_{\Delta_2},
\end{equation}
with
\begin{align}
H_{\mathrm{ABA},0}&=\left(\begin{array}{cccccc}
 0 &v_0\pi^\dagger &v_4\pi^\dagger &v_3\pi &\frac{1}{2}\gamma_2 &0  \\
v_0\pi &\delta &\gamma_1 &v_4\pi^\dagger &0 &\frac{1}{2}\gamma_5 \\ 
v_4\pi & \gamma_1 &\delta &v_0\pi^\dagger &v_4\pi &\gamma_1 \\
v_3\pi^\dagger &v_4\pi &v_0\pi &0 &v_3\pi^\dagger &v_4\pi  \\
\frac{1}{2}\gamma_2 &0 &v_4\pi^\dagger &v_3\pi &0 &v_0\pi^\dagger  \\
0 &\frac{1}{2}\gamma_5 &\gamma_1 &v_4\pi^\dagger &v_0\pi &\delta
\end{array}\right),\\ 
H_{\mathrm{ABC},0}&=\left(\begin{array}{cccccc}
 0 &v_0\pi^\dagger &v_4\pi^\dagger &v_3\pi &0 &\frac{1}{2}\gamma_2  \\
v_0\pi &\delta &\gamma_1 &v_4\pi^\dagger &0 &0  \\
v_4\pi & \gamma_1 &\delta &v_0\pi^\dagger &v_4\pi^\dagger &v_3\pi \\
v_3\pi^\dagger &v_4\pi &v_0\pi &\delta &\gamma_1 &v_4\pi^\dagger \\
0 &0 &v_4\pi &\gamma_1 &\delta &v_0\pi^\dagger \\
\frac{1}{2}\gamma_2  &0 &v_3\pi^\dagger &v_4\pi &v_0\pi &0
\end{array}\right),
 \\
H_{\Delta_1}&=\mathrm{diag}\left(\Delta_1,\Delta_1,0,0,-\Delta_1,-\Delta_1\right), \\ 
H_{\Delta_2}&=\mathrm{diag}\left(\Delta_2,\Delta_2,-2\Delta_2,-2\Delta_2,\Delta_2,\Delta_2\right),
\end{align}
where we have used $\left(A_1,B_1,A_2,B_2,A_3,B_3\right)$ as the basis. For this case $2\Delta_1$ is the potential difference between outer layers and $3\Delta_2$ corresponds to the difference between the mean potential of the outer layers and the middle layer. For trilayer systems we adopt the value $\Delta_2=-0.0023\,\mathrm{eV}$.

\subsection{Tetralayer graphenes}
There are three possible energetically stable stackings of tetralayer graphene, $\mathrm{st} = \mathrm{ABCA}$,  $\mathrm{ABAB}$, and $\mathrm{ABAC}$ (or equivalently $\mathrm{ABCB}$). Generally we can write the Hamiltonian in the form
\begin{equation}
H_\mathrm{st}=H_{0,st}+H_{\Delta_1}+H_{\Delta_2}+H_{\Delta_3},
\label{8times8Ham}
\end{equation}
where $H_{\Delta_1}$, $H_{\Delta_2}$ and $H_{\Delta_3}$ describe the electrostatic potentials on different layers and are independent of the stacking. In the basis $\left(A_1,B_1,A_2,B_2,A_3,B_3,A_4,B_4\right)$ they have the form
\begin{align}
H_{\Delta_1}&=\mathrm{diag}\left(\Delta_1,\Delta_1,\frac{\Delta_1}{3},\frac{\Delta_1}{3},-\frac{\Delta_1}{3},-\frac{\Delta_1}{3},-\Delta_1,-\Delta_1\right),\\
H_{\Delta_2}&=\mathrm{diag}\left(\Delta_2,\Delta_2,-\Delta_2,-\Delta_2,-\Delta_2,-\Delta_2,\Delta_2,\Delta_2\right), \\
H_{\Delta_3}&=\mathrm{diag}\left(0,0,-\Delta_3,-\Delta_3,\Delta_3,\Delta_3,0,0\right).
\end{align}
$H_{st,0}$ is the $8\times8$ Hamiltonian for each stacking case. They are written explicitly as
\begin{widetext}
\begin{align}
H_{\mathrm{ABCA},0}&=\left(\begin{array}{cccccccc}
 0 &v_0\pi^\dagger &v_4\pi^\dagger &v_3\pi &0 &\frac{1}{2}\gamma_2 &0 &0  \\
v_0\pi &\delta &\gamma_1 &v_4\pi^\dagger &0 &0 &0 &0 \\
v_4\pi & \gamma_1 &\delta &v_0\pi^\dagger &v_4\pi^\dagger &v_3\pi &0 &\frac{1}{2}\gamma_2 \\
v_3\pi^\dagger &v_4\pi &v_0\pi &\delta &\gamma_1 &v_4\pi^\dagger &0 &0 \\
0 &0 &v_4\pi &\gamma_1 &\delta &v_0\pi^\dagger &v_4\pi^\dagger &v_3\pi \\
\frac{1}{2}\gamma_2  &0 &v_3\pi^\dagger &v_4\pi &v_0\pi &\delta &\gamma_1 &v_4\pi^\dagger \\
0 &0 &0 &0 &v_4\pi &\gamma_1 &\delta &v_0\pi^\dagger \\
0 &0 &\frac{1}{2}\gamma_2  &0 &v_3\pi^\dagger &v_4\pi &v_0\pi &0
\end{array}\right), \\
H_{\mathrm{ABAB},0}&=\left(\begin{array}{cccccccc}
 0 &v_0\pi^\dagger &v_4\pi^\dagger &v_3\pi &\frac{1}{2}\gamma_2 &0 &0 &0  \\
v_0\pi &\delta &\gamma_1 &v_4\pi^\dagger &0 &\frac{1}{2}\gamma_5 &0 &0 \\ v_4\pi & \gamma_1 &\delta &v_0\pi^\dagger &v_4\pi &\gamma_1 &\frac{1}{2}\gamma_5 &0 \\
v_3\pi^\dagger &v_4\pi &v_0\pi &0 &v_3\pi^\dagger &v_4\pi &0 &\frac{1}{2}\gamma_2 \\
\frac{1}{2}\gamma_2 &0 &v_4\pi^\dagger &v_3\pi &0 &v_0\pi^\dagger &v_4\pi^\dagger &v_3\pi \\
0 &\frac{1}{2}\gamma_5 &\gamma_1 &v_4\pi^\dagger &v_0\pi &\delta &\gamma_1 &v_4\pi^\dagger \\
0 &0 &\frac{1}{2}\gamma_5 &0 &v_4\pi &\gamma_1 &\delta &v_0\pi^\dagger \\
0 &0 &0 &\frac{1}{2}\gamma_2 &v_3\pi^\dagger &v_4\pi &v_0\pi &0
\end{array}\right), \\
H_{\mathrm{ABAC},0}&=\left(\begin{array}{cccccccc}
 0 &v_0\pi^\dagger &v_4\pi^\dagger &v_3\pi &\frac{1}{2}\gamma_2 &0 &0 &0  \\
v_0\pi &\delta &\gamma_1 &v_4\pi^\dagger &0 &\frac{1}{2}\gamma_5 &0 &0 \\ 
v_4\pi & \gamma_1 &\delta &v_0\pi^\dagger &v_4\pi &\gamma_1 &0 &0 \\
v_3\pi^\dagger &v_4\pi &v_0\pi &0 &v_3\pi^\dagger &v_4\pi &\frac{1}{2}\gamma_2 &0 \\
\frac{1}{2}\gamma_2 &0 &v_4\pi^\dagger &v_3\pi &\delta &v_0\pi^\dagger &v_4\pi &\gamma_1 \\
0 &\frac{1}{2}\gamma_5 &\gamma_1 &v_4\pi^\dagger &v_0\pi &\delta &v_3\pi^\dagger &v_4\pi \\
0 &0 &0 &\frac{1}{2}\gamma_2 &v_4\pi^\dagger &v_3\pi &0 &v_0\pi^\dagger \\
0 &0 &0 &0 &\gamma_1 &v_4\pi^\dagger &v_0\pi &\delta
\end{array}\right).
\end{align}
\end{widetext}
In all calculations we choose $\Delta_2=\Delta_3=0$, since those are supposed to be in a range of few meV (see the discussion of the screening below) and will not have important effect on the conclusions of the paper.
\begin{figure*}[t]
\centering
\includegraphics[width=.89\linewidth]{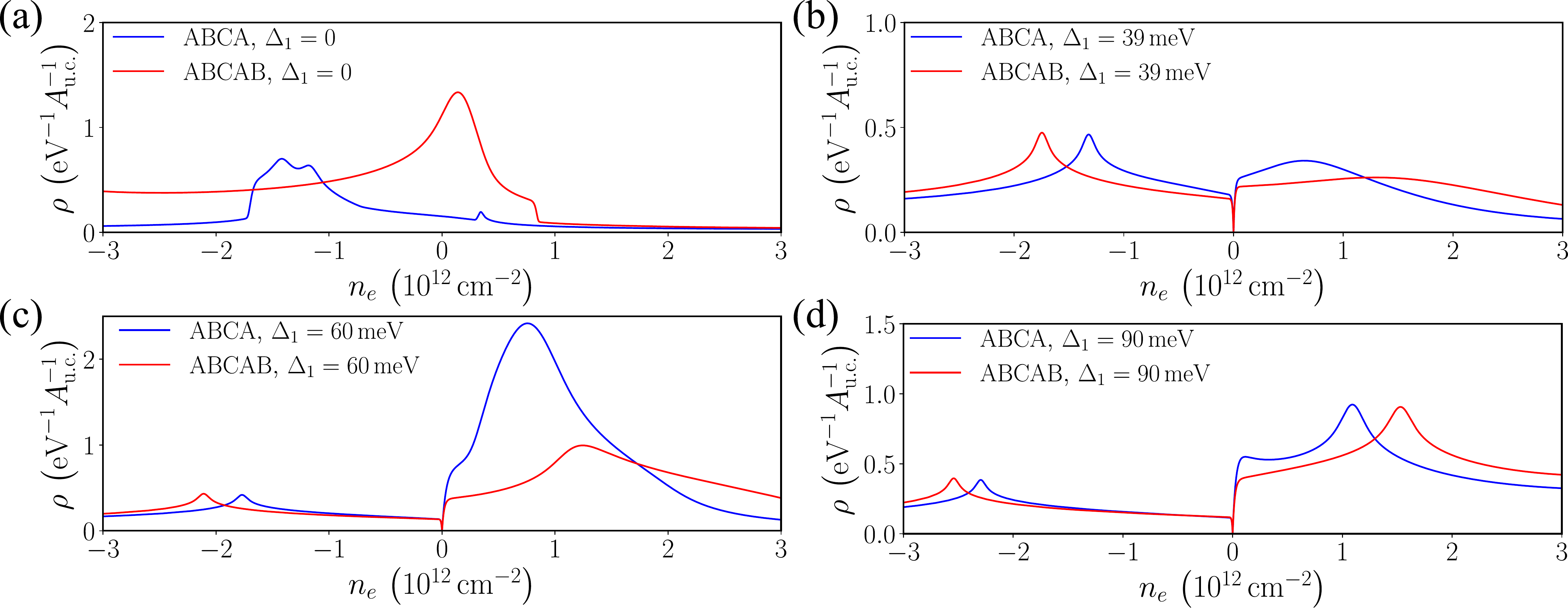}
\caption{Comparison of the DOS for tetralayer ABCA and pentalayer ABCAB for different values of the layer asymmetry potential. While for $\Delta_1=0$ ABCAB graphene has a higher DOS, at other values of the displacement field the DOS of ABCA is comparable or larger.
}

\label{fig:ABCACompareABCAB}
\end{figure*}
\subsection{Band structure of ABCAB graphene}
In Fig.~\ref{fig:ABCACompareABCAB} we compare the DOS for ABCA tetralayer with ABCAB pentalayer. While for $\Delta_1=0$ (generally for $\Delta_1<10\,\mathrm{meV}$) the additional layer gives some enhancement of the DOS, for larger values of $\Delta_1$, the tetralayer has a comparable or even larger DOS compared to the pentalayer. Due to the low lying additional bands in the latter, screening is expected to be stronger for the pentalayer than for the tetralayer, narrowing the experimentally accessible window in which the asymmetry potential can be tuned. Therefore, adding more layers ceases to be helpful beyond four layers, which is why we concentrate on systems with up to four layers in the main text.

\subsection{Band structure of ABAC/ABCB graphene}
\begin{figure*}[t]
\centering
\includegraphics[width=.89\linewidth]{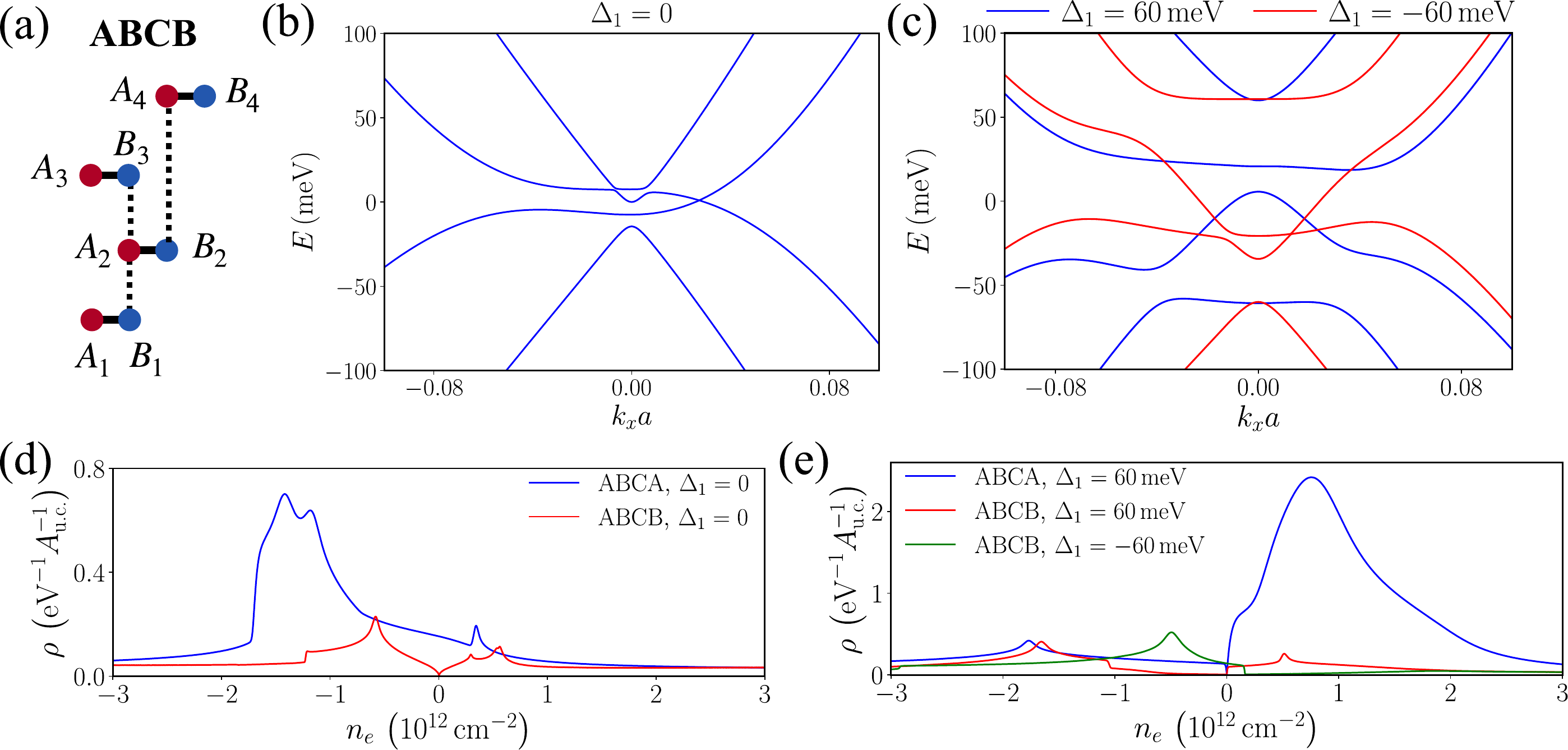}
\caption{Stacking order (a) and energy dispersion (b,c) for ABCB graphene with zero and non-zero displacement field $\Delta_1$.  (d,e) Comparison of the corresponding DOS of ABCB and ABCA stackings. Note the dependence of the band structure and DOS of ABCB graphene on the sign of $\Delta_1$.
}
\label{fig:ABCBBandStructure}
\end{figure*}

In the family of multilayer graphene stackings up to four layers, ABCB plays a special role since it both lacks an inversion center and mirror symmetry [cf. Fig.~\ref{fig:ABCBBandStructure}(a)]. Therefore, its band structure depends on the sign of the potential $\Delta_1$. This is shown in Fig.~\ref{fig:ABCBBandStructure} (b-c). For $\Delta_1=0$ the band structure is gapless and remains such also for negative $\Delta_1<0$, at least up to $\Delta_1=-60\,\mathrm{meV}$ [cf. Fig.~\ref{fig:ABCBBandStructure} (c)]. It should be mentioned that there is no symmetry protecting the band crossing for $\Delta_1<0$. 
Fig.~\ref{fig:ABCBBandStructure} (d,e) compares the DOS for ABCB and ABCA stackings. For the chosen tight-binding parameters, ABCA has a higher DOS for all values of the asymmetry potential. Therefore, we expect interaction effects and superconducting instabilities to be less pronounced in ABCB graphene compared to ABCA.

\begin{figure*}[t]
\centering
\includegraphics[width=.99\linewidth]{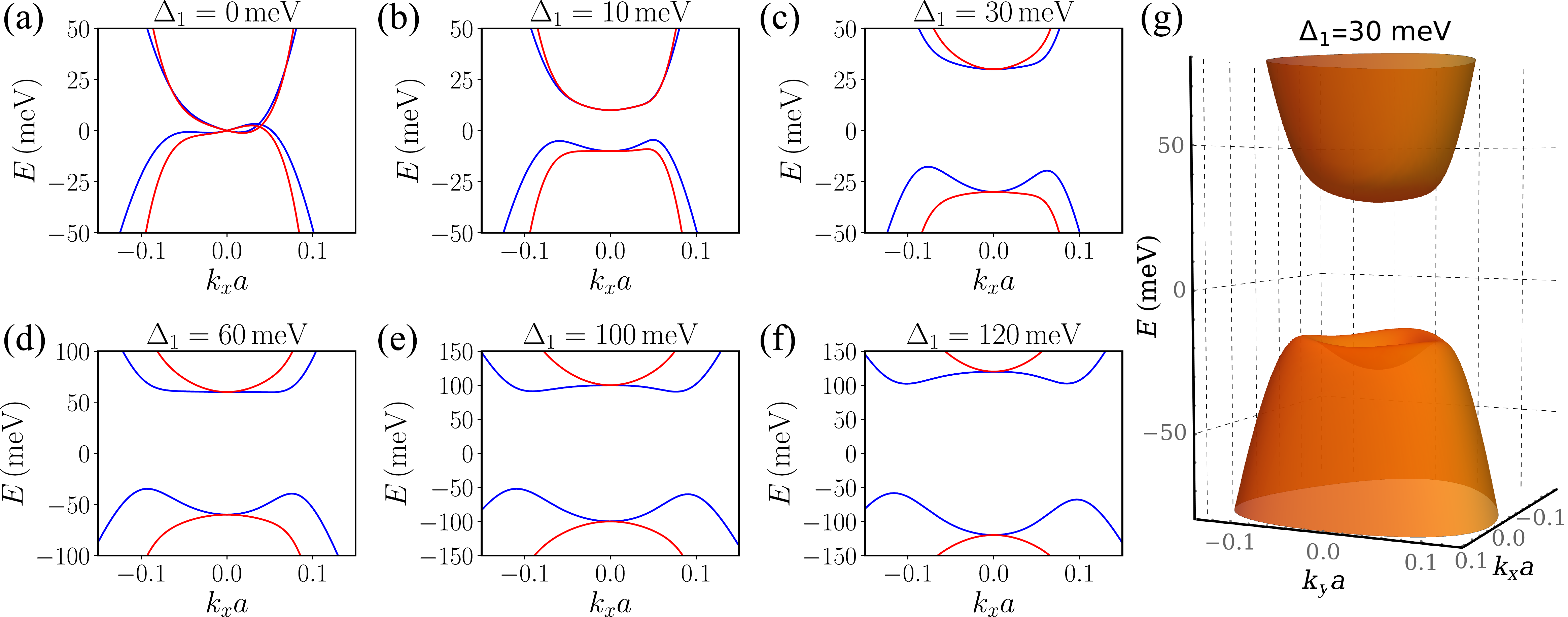}
\caption{
(a-f) Progression of the energy bands of ABCA graphene as a function of $k_xa$ with the change of the asymmetry potential $\Delta_1$. Blue (red) lines are from Hamiltonian \eqref{8times8Ham} (effective Hamiltonian \eqref{Eff2times2}). (g) 3D plot of the energy bands for $\Delta_1=30\,\mathrm{meV}$ from Hamiltonian \eqref{8times8Ham}.}
\label{fig:EnergyProgression}
\end{figure*}

\subsection{Low energy band structure of ABCA graphene}
Since in the remainder of our work we focus on ABCA stacking, we derive the low energy Hamiltonian corresponding to $A_1$ and $B_4$ sublattices. Following the standard procedure \cite{MacDonald2010} we get
\begin{widetext}
\begin{align}
h_\mathrm{eff}\left(\mathbf{k}\right)&=-\delta\frac{v^2_0k^2}{\gamma^2_1}\sigma_0+\Delta_2\sigma_0+\Delta_1\left(1+\frac{4v^2_0k^2}{3\gamma^2_1}\right)\sigma_z-\Delta_3\frac{v^2_0k^2}{\gamma^2_1}\sigma_z-\nonumber \\
&\underbrace{\frac{v^4_0k^4}{\gamma^3_1}\left(\cos\left(4\varphi_\mathbf{k}\right)\sigma_x+\xi\sin\left(4\varphi_\mathbf{k}\right)\sigma_y\right)}_\text{BP4}-
\frac{2v_0v_4k^2}{\gamma_1}\sigma_0-\underbrace{\frac{v^2_3k^2}{\gamma_{1}}\left(\cos\left(2\varphi_\mathbf{k}\right)\sigma_x-\xi\sin\left(2\varphi_\mathbf{k}\right)\sigma_y\right)}_\text{BP2}+ \nonumber \\
&\underbrace{\left(\frac{2v_0v_3v_4\xi}{\gamma^2_1}-\frac{v_0^2v_4\gamma_2\xi}{\gamma^3_1}\right)\left(k^3_x-3k_xk^2_y\right)\sigma_0}_\text{BP0}+\underbrace{\left(\frac{3v_0^2v_3k^3}{\gamma^2_1}-\frac{v^3_0\gamma_2k^3}{\gamma^3_1}-\frac{v_0\gamma_2k}{\gamma_1}\right)\left(\xi\cos\varphi_\mathbf{k}\sigma_x+\sin\varphi_\mathbf{k}\sigma_y\right)}_\text{BP1},
\label{Eff2times2}
\end{align}
\end{widetext}
where we assumed $\gamma_2\ll\gamma_1$ and kept only $v_3/\gamma_1$ and $v_4/\gamma_1$ terms up to first order.

Fig.~\ref{fig:EnergyProgression} shows the progression of the energy levels with the change of asymmetry potential $\Delta_1$ and comparing the results from $8\times8$ Hamiltonian \eqref{8times8Ham} and from effective Hamiltonian \eqref{Eff2times2}. As was shown in the main text, a non-zero asymmetry potential opens a gap. For small values of $\Delta_1$ the electron band has a single Fermi surface and for those cases the effective Hamiltonian \eqref{Eff2times2} captures the main features quite well. For higher values of $\Delta_1$ and at $n_e>-3\times10^{12}\,\mathrm{cm^{-2}}$ charge densities of the hole band the Fermi surface consists of three pockets or an annulus. For those cases the effective $2\times2$ model is less useful since it is an expansion of $8\times8$ Hamiltonian that relies on $k$ being a small parameter. 

\begin{figure*}[t]
\centering
\includegraphics[width=.9\linewidth]{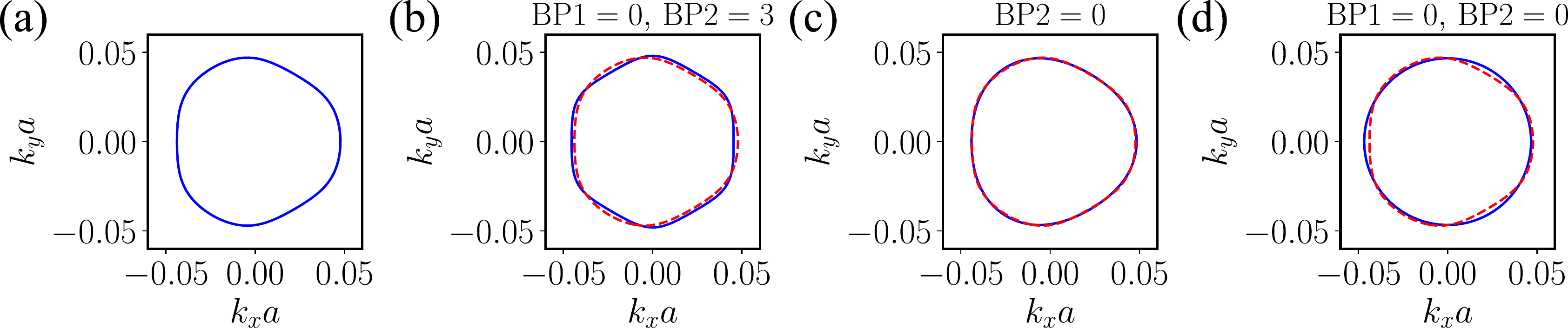}
\caption{
Fermi surfaces of ABCA graphene for $\Delta_1=9\,\mathrm{meV}$ and $n=1.1\times10^{12}\,\mathrm{cm^{-2}}$ obtained from the effective Hamiltonian \eqref{Eff2times2} is shown in (a). All subsequent panels illustrate the change of the FS when specific terms in effective Hamiltonian are set to zero. In (b) $\mathrm{BP2}=3$ title means the BP2 term was magnified three times to make the effect of the term more pronounced. When $\mathrm{BP1}=0$, $\mathrm{BP2}$ causes sixfold warping of the Fermi surface. In contrast $\mathrm{BP1}$ causes trigonal warping (c). When both $\mathrm{BP1}=0$ and $\mathrm{BP2}=0$ the Fermi surface is circular (d). In (b-d) red dashed lines show the FS from panel (a) for comparison.}
\label{fig:FSEffective}
\end{figure*}

In the Hamiltonian \eqref{Eff2times2} we observe the presence of the terms with Berry phase equal to $4$, $2$, and $1$ denoted as BP4, BP2, and BP1, respectively. The last line of \eqref{Eff2times2}  also contains a term (denoted as BP0) which causes additional trigonal warping of the Fermi surface and is proportional to the identity matrix in sublattice space, $\sigma_0$. For the current values of tight-binding parameters this term gives a small contribution and we do not consider it further. Fig.~\ref{fig:FSEffective} shows the change of the Fermi surface of the electronic band as  different terms in \eqref{Eff2times2} are set to zero. For the case of $\mathrm{BP1}=0$ and $\mathrm{BP2}=0$ the Fermi surface becomes circular. A nonzero BP2 term results in a \emph{sixfold} warping of the surface, so that it becomes reminiscent of a hexagon shape. 
Finally the term with Berry phase equal to one (BP1) leads to trigonal warping. Since BP1 contains contributions of order $k$ and $k^3$ with opposite signs, at a certain value of the density these two contributions cancel each other. Therefore, upon changing the carrier density it is possible to flip the \emph{orientation} of the trigonal warping.  To the best of our knowledge, this is the unique feature of the tetralayer that does not appear in mono- bi- or trilayer graphenes.

\subsection{Screening in tetralayer ABCA graphene}

\begin{figure*}[t]
\centering
\includegraphics[width=.9\linewidth]{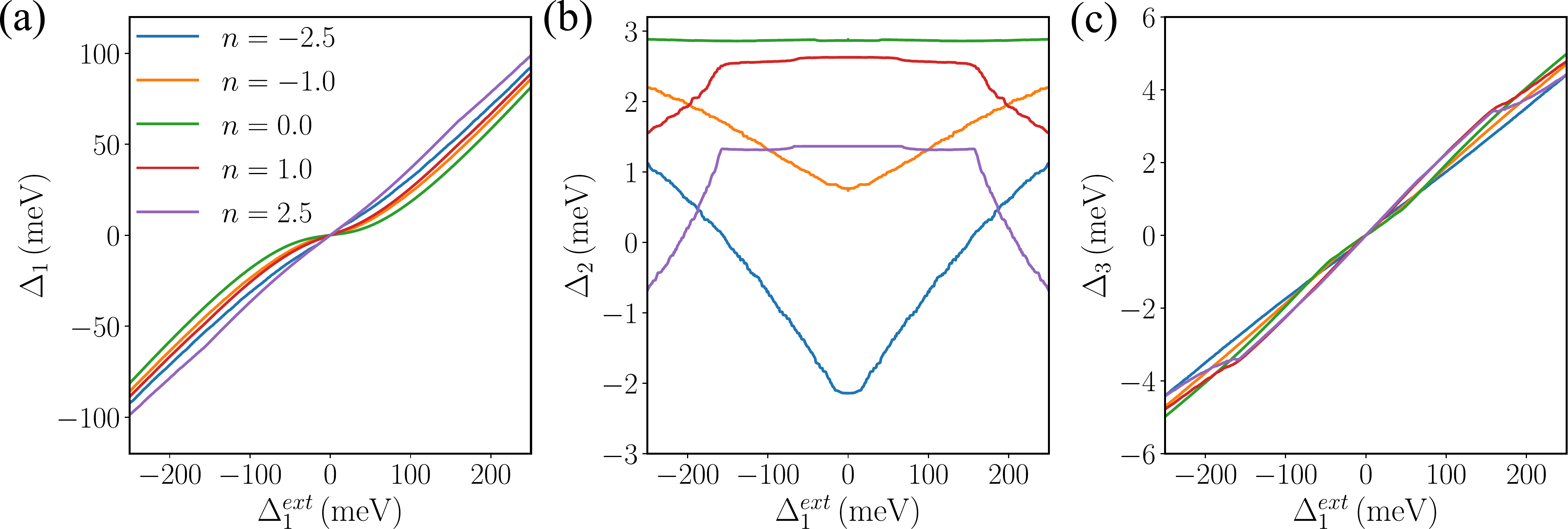}
\caption{
Dependence of $\Delta_1$ (a), $\Delta_2$ (b), and $\Delta_3$ (c) parameters for ABCA graphene on external layer asymmetry potential $\Delta_{1,ext}$ determined from self-consistent Hartree screening. The calculation employs $1000\times1000$ grid points from full Brillouin zone. Densities are given in units of $10^{12}\,\mathrm{cm^{-2}}$.}
\label{fig:Screening}
\end{figure*}
To determine the experimentally accessible range of $\Delta_1$ and the values of $\Delta_2$ and $\Delta_3$, we consider the self-consistent Hartree screening due to the presence of gates \cite{Koshino2009gate}. Denoting the layer potentials as $u_i$, we define the $\Delta_i$ parameters for tetralayer graphene stacks as
\begin{align}
\Delta_1&=\frac{u_1-u_4}{2}, \\
\Delta_2&=\frac{u_1-u_2-u_3+u_4}{4}, \\
\Delta_3&=\frac{u_1-3u_2+3u_3-u_4}{6}.
\end{align}
Assuming that average potential in the graphene multilayer is zero yields the constraint $u_1+u_2+u_3+u_4=0$. Then, using standard electrostatics we can relate $\Delta_i$ to the potential on the gates and the charge densities $n_i$ at each layer, 
\begin{align}
\Delta_1&=\frac{3}{4}\frac{ed}{\epsilon_r}\left(\frac{\epsilon_tV_t}{L_t}-\frac{\epsilon_bV_b}{L_b}\right)+\frac{e^2d}{4\epsilon_r}\left[3\left(n_1-n_4\right)+\left(n_2-n_3\right)\right], \\
\Delta_2&=-\frac{e^2d}{4\epsilon_r}\left(n_2+n_3\right), \\
\Delta_3&=-\frac{e^2d}{6\epsilon_r}\left(n_2-n_3\right).
\end{align}
Here $d=3.34\,\AA$ is the distance between the layers, while $V_t$ ($V_b$) is the potential on top (bottom) gate, $L_t$ ($L_b$) are the distances between top (bottom) gate and the graphene sample, while likewise $\epsilon_t$ ($\epsilon_b$) are the dielectric constants between top (bottom) gate and the graphene sample. Identifying $\Delta_{1,\text{ext}}=\frac{3}{4}\frac{ed}{\epsilon_r}\left(\frac{\epsilon_tV_t}{L_t}-\frac{\epsilon_bV_b}{L_b}\right)$ as  the external potential we arrive at a system of equations that can be solved self-consistently, yielding an estimate of the external potentials accessible in experiment. For concreteness, we fix the dielectric constant as $\epsilon_r=2$. The displacement field is defined as $D=\frac{\epsilon_tV_t}{L_t}-\frac{\epsilon_bV_b}{L_b}$ \cite{Zibrov2018emergent}. In experiment, the upper limit of $D$ is typically in the order of $D_\text{max}=2.0\,\mathrm{V/nm}$, which yields $\Delta^\text{max}_{1,\text{ext}}\approx250\,\mathrm{meV}$. 

Fig.~\ref{fig:Screening} shows the dependence of the $\Delta_1$, $\Delta_2$ and $\Delta_3$ potentials on external potential $\Delta_{1,ext}$ as calculated from self-consistent Hartree screening. Note that this calculation requires knowledge of the charge density distribution between layers. This has a contribution from all momenta in the Brillouin zone and therefore, a full zone sampling is necessary. As can be seen from the figure, $\Delta_2$ and $\Delta_3$ are in the range of a few meV and can be safely set to zero, since that range does not have a noticeable effect on the band structure. Taking into account the effect of the screening and using the maximum value of the external potential, $\Delta_1$ can take values up to $100-120$ meV, which is exceeding any of the displacement field parameters discussed in the main text by a fair margin. Therefore, we expect that all the effects described in this work can be observed with presently available methods.    

\section{Fermiology of ABCA graphene}

\begin{figure}[t]
\centering
\includegraphics[width=.99\linewidth]{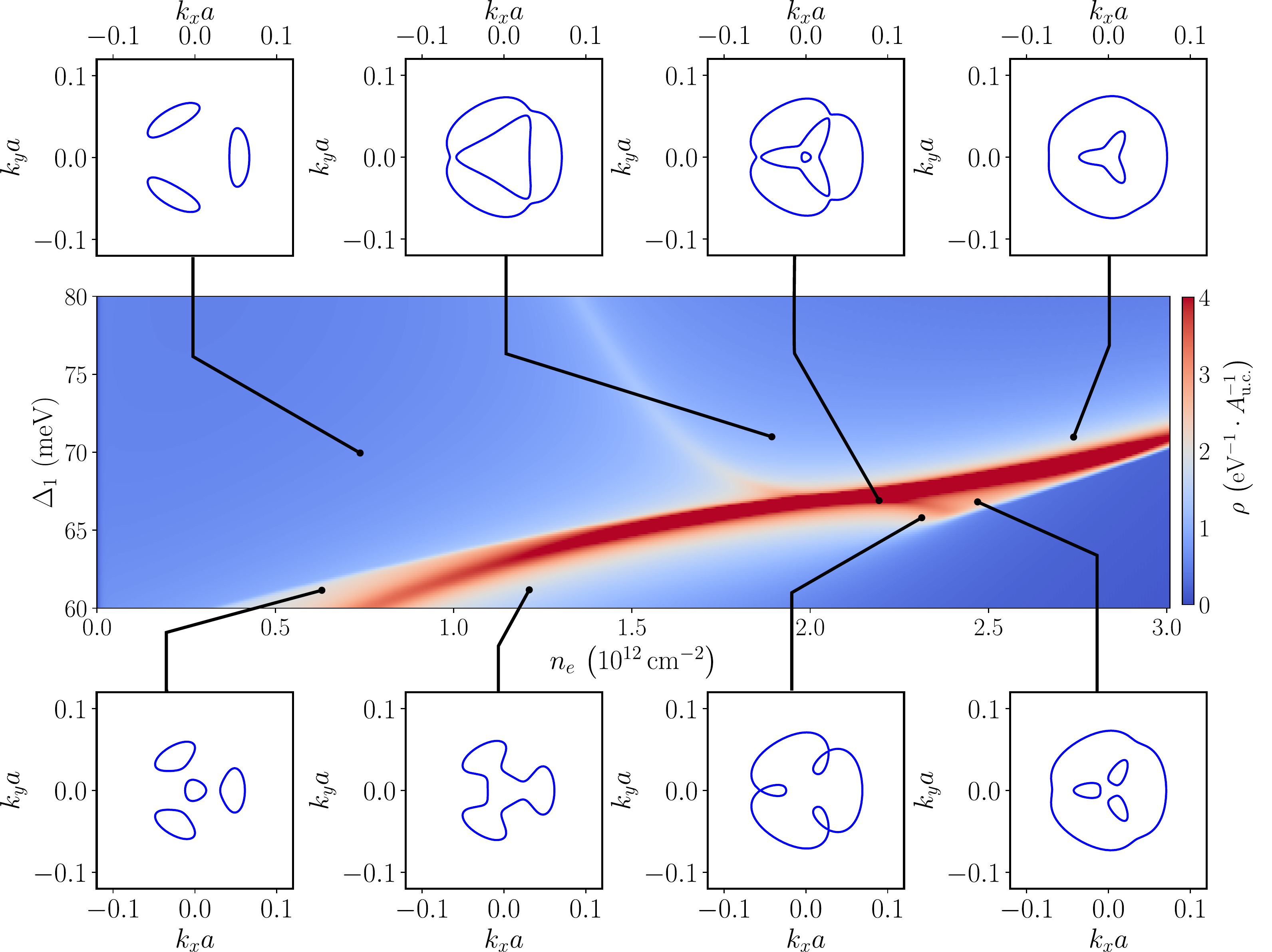}
\caption{Color plot of the DOS $\rho$ for the ABCA tetralayer, similar to Fig.~2 in the main text but zoomed in for large displacement field on the electron side. In this region of the phase diagram several VHS occur in close proximity. Representative Fermi surfaces are shown for selected locations, some of which appear only close to these Lifshitz points and not anywhere else in the phase diagram.}
\label{fig:DOSFSZoom}
\end{figure} 

In this section we discuss the high DOS region on the electron side where several Van Hove singularities (VHS) meet. This leads to a higher order VHS, similar to the one discussed in the context of twisted bilayer graphene \cite{Yuan2019}. The resulting DOS as a function of $\Delta_1$ and $n$ is shown in Fig.~\ref{fig:DOSFSZoom}, alongside with a few representative FS which emerge. The fermiology is very rich in a narrow region of densities and displacement fields in the vicinity of the higher order VHS. At the higher order VHS, where 
five conventional VHS nearly collide, we observe three FS originating from the same band. There are also regions with four pockets with different topology. The divergence of the DOS is higher than at the individual VHS and interaction effects are expected to be dominant. In the Stoner interaction model considered in the main text (cf. next section) the system gets strongly polarized and such FS are only observed in the region of the phase diagram with full spin and valley polarization (1x degenerate).   

\section{Stoner transitions}

\begin{figure}[b]
\centering
\includegraphics[width=.99\linewidth]{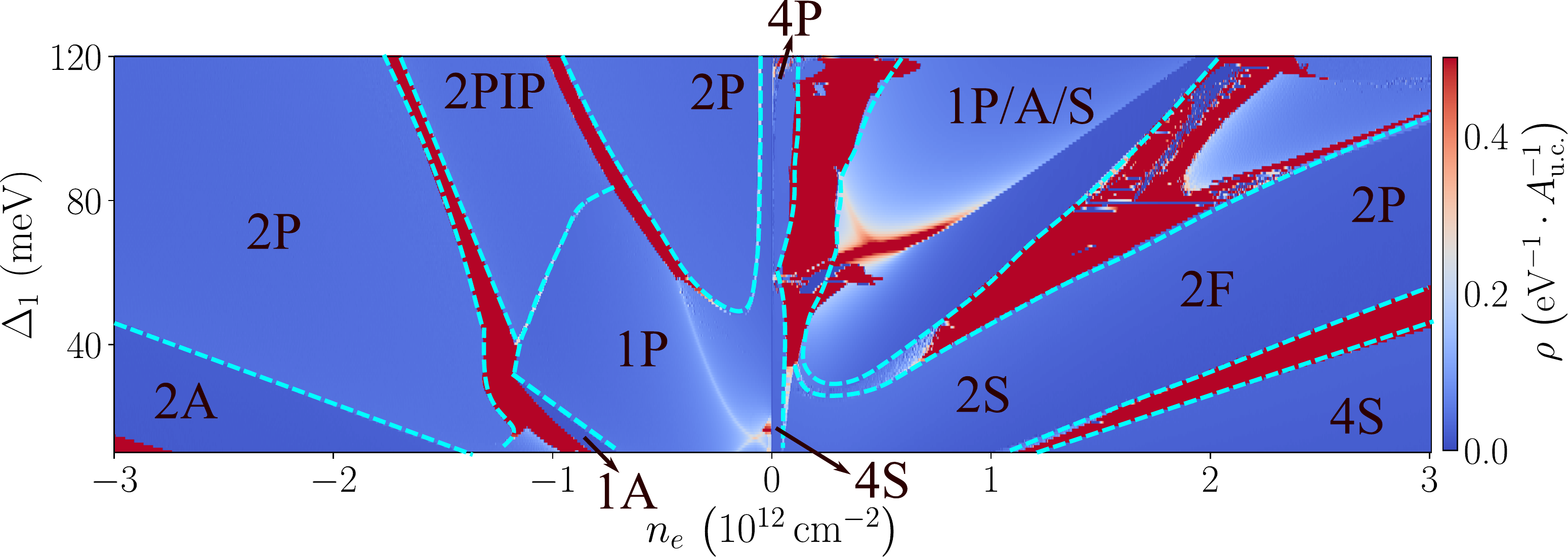}
\caption{Stoner phase diagram for interaction strengths $U=30\,\mathrm{eV}$ and $J=-9\,\mathrm{eV}$. The labels are the same as in the Stoner diagram (Fig.~3) of the main text.}
\label{fig:StonerSup}
\end{figure} 

As for the case of trilayer graphene \cite{Zhou2021half}, Stoner transitions are obtained through the minimization of the grand potential density
\begin{equation}
\frac{\Phi}{A}=\sum_\alpha E_{0}\left(n_\alpha\right)+V_\mathrm{int}-\mu\sum_\alpha n_\alpha,    
\end{equation}
where $E_0$ is the kinetic energy and $n_\alpha$ is density of each flavor corresponding to spin and valley. $V_\mathrm{int}$ is the interaction potential which besides a SU(4) symmetric term also includes the scattering between valleys, namely
\begin{equation}
V_\mathrm{int}=\frac{UA_\mathrm{u.c.}}{2} \sum_{\alpha\neq\beta}n_\alpha n_\beta+JA_\mathrm{u.c.}\left(n_1-n_3\right)\left(n_2-n_4\right),
\end{equation}
where we used the indexing $1=\{K,\uparrow\}$, $2=\{K^\prime,\uparrow\}$, $3=\{K,\downarrow\}$, $4=\{K^\prime,\downarrow\}$ and $A_\mathrm{u.c.}=\sqrt{3}a^2/2$ is the area of unit cell. Here $U$ and $J$ are the interaction constants at the unit cell level. For ABC trilayer parameter values $U=30\,\mathrm{eV}$ and $J=-9\,\mathrm{eV}$ has been used \cite{Zhou2021half}. We show the Stoner phase diagram for ABCA tetralayer in Fig.~\ref{fig:StonerSup}. Due to the larger number of layers we expect the strength of $U$ and $J$ be weaker for tetralayer compared to trilayer graphene. Therefore, in the main text we show Stoner phase diagram for $U=15\,\mathrm{eV}$ and $J=-4.5\,\mathrm{eV}$. Comparing the two diagrams we see qualitatively similar phases. The triple degenerate phase is absent in both cases. For weaker interaction strength the so-called partially isospin polarized (PIP) phases that are characterized by unequal population of different flavors are more prevalent. More importantly, for weaker values of the interaction, the fourfold degenerate phase on the electron side spans a broader range of densities and asymmetry potential, thereby also pushing the double degenerate phase to higher $\Delta_1$ and lower $n$. 
Since the electronic mechanism for superconductivity which we consider in this work only yields topological superconductivity from a double degenerate parent phase, weaker values of the interaction are therefore more favorable towards realizing topological superconductivity.

\section{Leading superconducting instabilities}
We investigate the superconducting instabilities driven by the long-range Coulomb interaction. To this end, the effective interaction between electrons is calculated incorporating particle-hole fluctuations through the random phase approximation (RPA) \cite{Ghazaryan2021unconventional}. The interaction potential is then given by
\begin{equation}
V_{\bm{q}}=\frac{V_{0,\bm{q}}}{1+\Pi_{0,\bm{q}}V_{0,\bm{q}}},    
\end{equation}
and the effective interaction is treated as instantaneous. Here $V_{0,\bm{q}}=\frac{2\pi e^{2}}{\epsilon q}\tanh(qd)$ is the screened Coulomb interaction and 
$$\Pi_{0,\bm{q}}=N\sum_{\bm{k}}\left|\Lambda_{\bm{k},\bm{q},\tau}\right|^{2}\frac{f(\varepsilon_{\bm{k},\tau})-f(\varepsilon_{\bm{k}+\bm{q},\tau})}{\varepsilon_{\bm{k}+\bm{q},\tau}-\varepsilon_{\bm{k},\tau}}$$ is the static polarization function, where $N$ defines flavour degeneracy, $\varepsilon_{\bm{k},\tau}$ is the energy of the electron in valley $\tau$, $f(x)$ is the Fermi-Dirac distribution and $\Lambda_{\bm{k},\bm{q},\tau}=\langle u_{\bm{k},\tau}\vert u_{\bm{k}+\bm{q},\tau}\rangle$ is the overlap matrix element between states of the electron band at momenta $\bm{k}$ and $\bm{k}+\bm{q}$. Given the interaction potential $V_{\bm{q}}$, the superconducting instability can be determined by solving the linearized BCS gap equation with a linear operator
\begin{equation}
[\mathcal{M}\Delta]_{\bm{k}}=-\int\frac{dk'_{\parallel}}{(2\pi)^{2}v_{\bm{k}'}}V_{\bm{k}-\bm{k}'} |\Lambda_{\bm{k},\bm{k}'-\bm{k},+1}|^2 \Delta_{\bm{k}'},
\label{linGap}
\end{equation}
where the integral is projected onto Fermi surface, $v_{\bm{k}}$ is the Fermi velocity and $\Delta_{\bm{k}}$ is the order parameter to be determined. The calculation proceeds by discretizing \eqref{linGap} and solving the eigenvalue matrix equation ${\mathcal{M}}_{\bm{k},\bm{k^\prime}}\phi_{\bm{k^\prime}}=\lambda\phi_{\bm{k}}$ \cite{Ghazaryan2021unconventional}.\\[-32pt]

\section{Topological superconductivity}

\begin{figure}[tb]
\centering
\includegraphics[width=.95\linewidth]{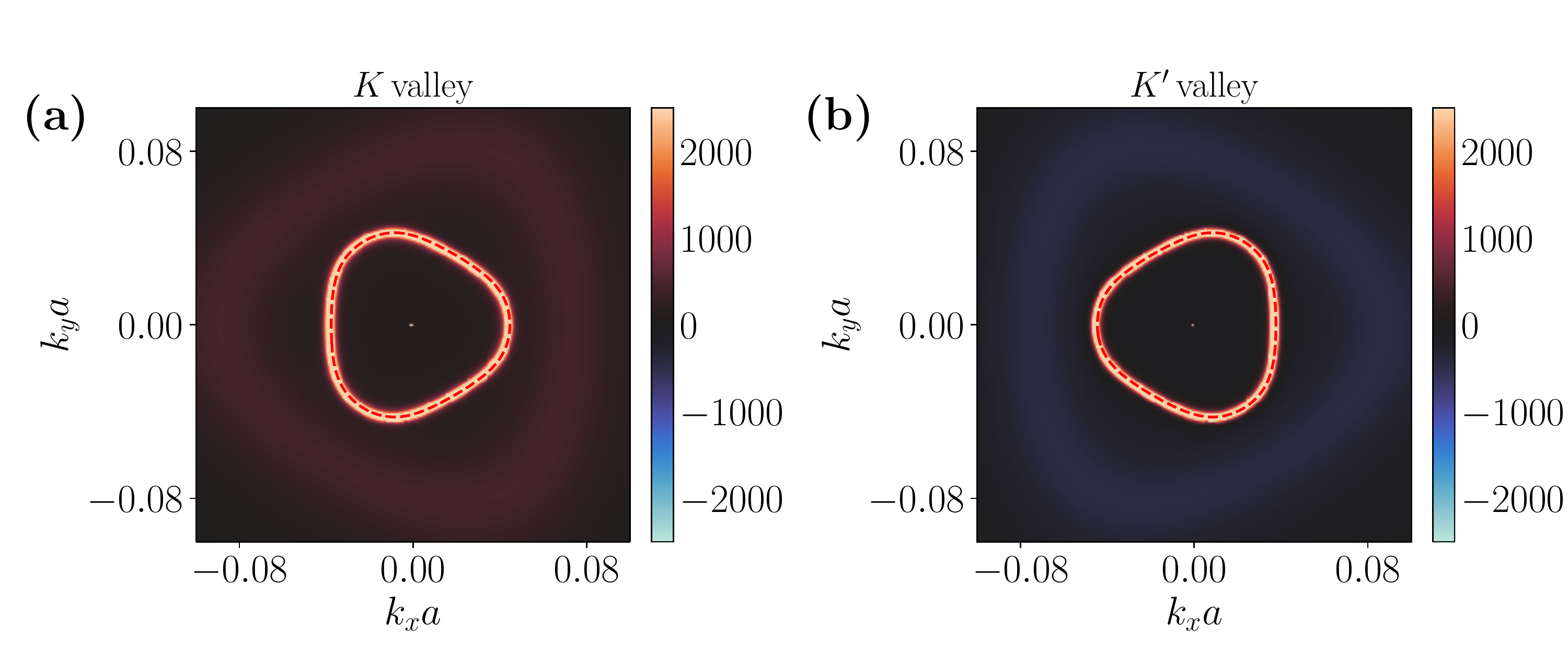}\\
\includegraphics[width=.95\linewidth]{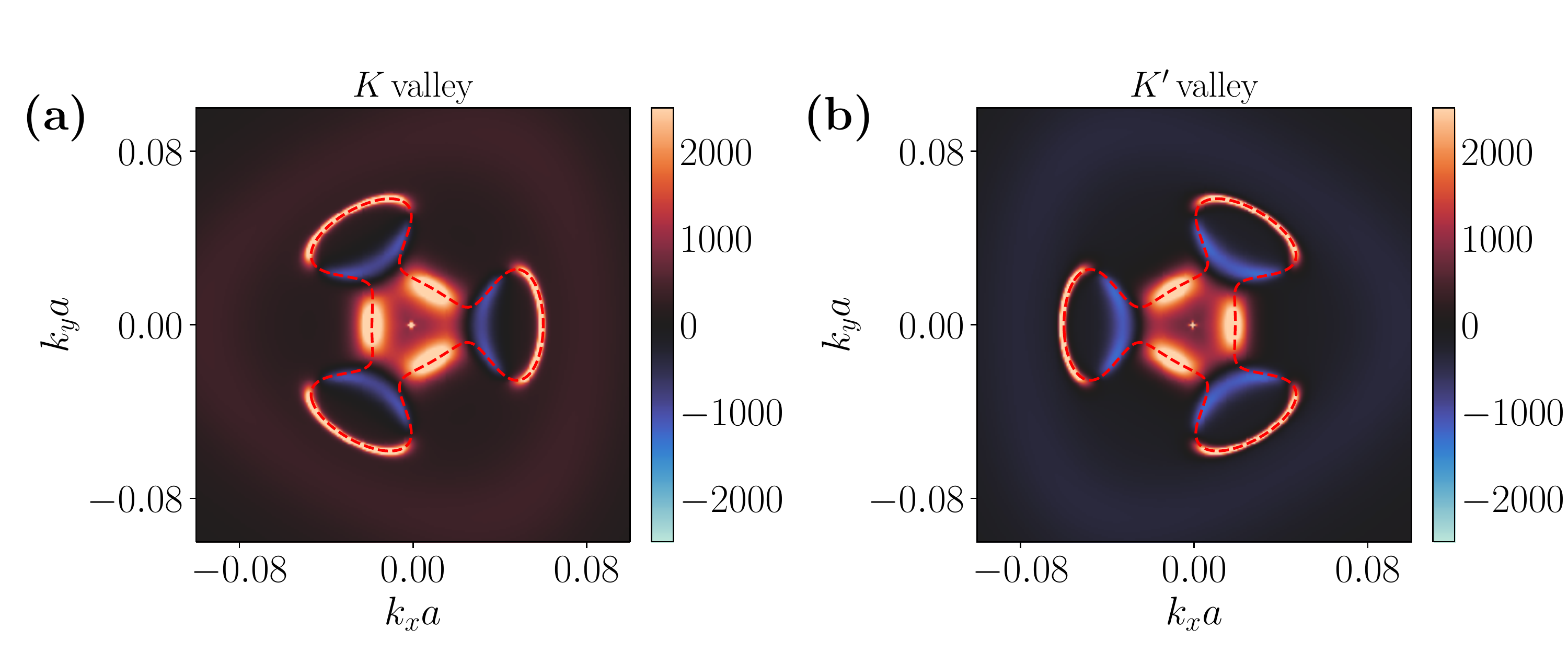}\\
\includegraphics[width=.95\linewidth]{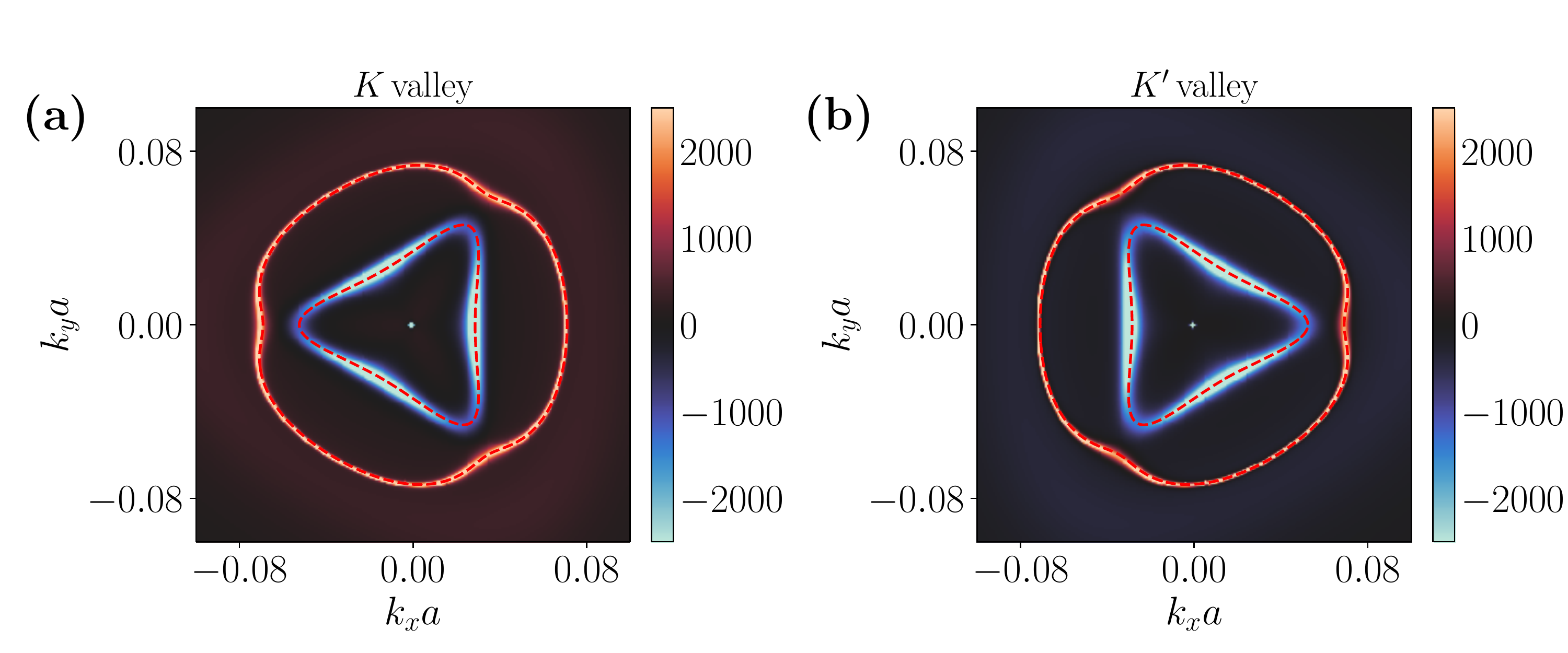}
\caption{
(top) Berry curvature obtained from BdG equations for ABCA stacking assuming $p_x+ip_y$ pairing in the electronic band for $\Delta_1=42\,\mathrm{meV}$, $\mu=43.4\,\mathrm{meV}$ and $\Delta_p=0.1\,\mathrm{meV}$. Red dashed lines show the Fermi surface of the parent Hamiltonian. The Chern number due to the superconducting pairing is 1 for both valleys.
(middle) The same as in top panel but for $\Delta_1=60\,\mathrm{meV}$ and $\mu=60.08\,\mathrm{meV}$. The Chern number due to the superconducting pairing is 1 for both valleys.
(bottom)
The same as in top panels but for $\Delta_1=69\,\mathrm{meV}$ and $\mu=68.77\,\mathrm{meV}$. The Chern number due to the superconducting pairing is 0 for both valleys.
} 
\label{fig:BerryCurvatureS}
\end{figure}
As is noted in the main text, for some range of parameters we obtain a superconducting instability with $p$-wave symmetry from the solution of the linearized gap equation. Based on general thermodynamic arguments~\cite{Ghazaryan2021unconventional} one can conclude that a system with $C_3$ symmetry favors chiral $p_x+ip_y$ type pairing below $T_c$. This reasoning applies directly to the case of the ABCA tetralayer, which has $C_3$ symmetry. This raises the interesting possibility that the resulting phase may be a topological superconductor. We address this question by calculating the Chern number due to the superconducting pairing for three representative cases of the Fermi surface topology.

To answer whether topological superconductivity can emerge in ABCA tetralayer graphene, we calculate the Chern number of the superconducting state. To this end, we construct the Bogoliubov–de Gennes (BdG) Hamiltonian for the order parameter $\Delta_{\bm{k}}=\Delta_p \left(k_x+ik_y\right)/k$. Once the energies and wave functions of the BdG equation are determined, the Berry curvature is calculated in a discretized Brillouin zone through the method of link variables \cite{fukui2005chern}. Fig.~\ref{fig:BerryCurvatureS} shows the Berry curvature for three different Fermi surfaces for which the stability analysis yielded $p$-wave pairing. Integrating the Berry curvature near $K$ and $K^\prime$ valleys shows that the Chern number due to the superconducting pairing is 1 for each valley when the Fermi surface consists of a single pocket. Therefore both the top and middle panels in Fig.~\ref{fig:BerryCurvatureS} correspond to topological superconductors with a total Chern number equal 2. In contrast, for an annular Fermi surface the Berry curvature near two Fermi surface contours has opposite sign and the overall Chern number is zero (cf. bottom panel in Fig.~\ref{fig:BerryCurvatureS}). Therefore, for an annular Fermi surface the resulting superconducting state is topologically trivial despite having a chiral $p_x+ip_y$ order parameter.

\end{document}